\documentclass[fleqn,10pt]{wlscirep}
\usepackage[utf8]{inputenc}
\usepackage[T1]{fontenc}
\usepackage{amssymb,amsthm,amsmath,bbm,bbold,graphicx,epsfig,threeparttable,color}%%{\color{red} text}
\usepackage[normalem]{ulem} %畫下底線封包,標籤
\usepackage{enumerate}
\usepackage{enumitem}
\usepackage[title]{appendix}
\usepackage{hyperref}
\usepackage{lineno}%加line numeber

\usepackage{upgreek}
\theoremstyle{definition}
\usepackage{braket}
\usepackage{xcolor}
\usepackage{bm}

\usepackage{url}
\newcommand{\proj}[1]{\ket{#1}\bra{#1}}
\newcommand{\ts}{\otimes}
\newcommand{\id}{\mathbb{1}}
\newcommand{\Tr}{\text{Tr}}
\newcommand{\TrA}{\text{Tr}_{A}}

\newcommand{\Cd}{C}
\newcommand{\Hd}{H_{\Delta}}

\newcommand{\CCd}{C^{\star}}
\newcommand{\CH}{H^{\star}_{\Delta}}

\newcommand{\assemb}[1]{\left\{#1\right\}}

\newcommand{\blue}[1]{{\color{blue} #1}}

\newcommand{\comment}[1]{}

\theoremstyle{definition}
\newtheorem{defi}{Definition}%[section]
\newtheorem{prop}[defi]{Property}

\title{Unveiling quantum steering by quantum-classical uncertainty complementarity}

\author[1,2,*]{Kuan-Yi Lee}
\author[1,2,*]{Jhen-Dong Lin}
\author[3,$\dagger$]{Karel Lemr}
\author[4]{Anton\'{i}n \v{C}ernoch}
\author[5,6]{Adam Miranowicz}
\author[5,7,8]{Franco Nori}
\author[9,$\ddagger$]{Huan-Yu Ku}
\author[1,2,5,10,$\S$]{Yueh-Nan Chen}
\affil[1]{Center for Quantum Frontiers of Research and Technology (QFort), National Cheng Kung University, Tainan 701, Taiwan}
\affil[2]{Department of Physics, National Cheng Kung University, Tainan 701, Taiwan}
\affil[3]{Palack\'{y} University in Olomouc, Faculty of Science, Joint Laboratory of Optics of Palack\'{y} University and Institute of Physics AS CR, 17. listopadu 12, 771 46 Olomouc, Czech Republic}
\affil[4]{Institute of Physics of the Academy of Sciences of the Czech Republic,
Joint Laboratory of Optics of Palack\'{y} University and Institute of
Physics AS CR, 17. listopadu 50a, 772 07 Olomouc, Czech Republic}
\affil[5]{Theoretical Quantum Physics Laboratory, Cluster for Pioneering Research, RIKEN, Wakoshi, Saitama 351-0198, Japan}
\affil[6]{Institute of Spintronics and Quantum Information, Faculty of Physics and Astronomy, Adam Mickiewicz University, 61-614 Poznań, Poland}
\affil[7]{Center for Quantum Computing, RIKEN, Wakoshi, Saitama 351-0198, Japan}
\affil[8]{Department of Physics, The University of Michigan, Ann Arbor, 48109-1040 Michigan, USA}
\affil[9]{Department of Physics, National Taiwan Normal University, Taipei 11677, Taiwan}
\affil[10]{Physics Division, National Center for Theoretical Sciences, Taipei 106319, Taiwan}
\affil[*]{These authors contributed equally}
\affil[$\dagger$]{k.lemr@upol.cz}
\affil[$\ddagger$]{huan.yu@ntnu.edu.tw}
\affil[$\S$]{yuehnan@mail.ncku.edu.tw}

\begin{abstract}
One of the remarkable aspects of quantum steering is its ability to violate local uncertainty complementarity relations. 
In this vein of study, various steering witnesses have been developed. 
Here, we introduce a novel complementarity relation between system's quantum and classical uncertainties corresponding to the distillable coherence and the von-Neumann entropy, respectively. 
We show that the proposed complementarity relation is tighter than the entropic uncertainty relation (EUR).
Leveraging this result, we propose a steering witness that is more efficient than the EUR. 
From the operational perspective, the steering witness quantifies the amount of extra distillable coherence facilitated by quantum steerability. 
Notably, the proposed steering witness serves as a full entanglement measure for pure bipartite states--an ability that the EUR lacks.
We also experimentally validate such a property through a photonic system. 
Furthermore, a deeper connection to the uncertainty principle is revealed by showcasing the steering-induced distillable coherence can quantifies measurement incompatibility and quantum steerability under genuine incoherent operations. 
Our work establishes a clear quantitative and operational link between coherence and steering, which are vital resources of quantum technologies, and underscores our efforts in bridging the uncertainty principle with quantum coherence.
\end{abstract}

\begin{document}

\flushbottom
\maketitle
\section*{Introduction}

Quantum steering~\cite{Wiseman2007PRL}, as a type of quantum correlations that is classified between Bell nonlocality and quantum entanglement, has garnered significant attention due to its applications in one-sided device-independent quantum information processing~\cite{PhysRevLett.107.170403,Piani2015,Li2015,Huang2019,JDLin2021PRA,Huang2021,Ku2022PRX,ku2023pra}, including quantum random number generation~\cite{Skrzypczyk2018,Guo2019}, quantum key distribution~\cite{Branciard2012,Bartkiewicz2016PRA}, quantum metrology~\cite{Yadin2021NC,Lee2023PRR}, and thermodynamics~\cite{Hsieh2024-1,Hsieh2024-2} (see also recent reviews~\cite{Cavalcanti_2017,Uola2020RMP,Xiang2022PRXQuantum}). 
In addition, it exhibits a profound connection to the local uncertainty principle. 
In its original formulation, known as Reid's criteria~\cite{Reid1989,RMP2009:Reid}, quantum steering is characterized by violating the Robertson-Schr\"{o}dinger uncertainty relation. To date, the notion of characterizing quantum steering by its ability to violate uncertainty relation has been generalized to different forms, including the entropic uncertainty relation (EUR)~\cite{Deutsch1983,Maassen1988,Fuchs1996,Wehner2010,Coles2017Rmp,KARTHIK2018635}, the complementarity of coherence for mutually unbiased bases~\cite{Ivanovic1992,Mondal2017PRA,Hu2018PRA}, metrological complementarity~\cite{Yadin2021NC}, etc.
% \red{Furthermore, it has been shown that entanglement can lead to violations of the EUR~\cite{Berta2010}, which can also be extended to one-sided device-independent scenarios.}
Moreover, a resemblance between quantum steering and measurement incompatibility has been uncovered~\cite{Quintino2014PRL,Uola2014PRL,Uola2015PRL,Cavalcanti2016PRA,Zhao2020,Wang2023CommPhys}, further refining our understanding of the connection between steering and a generalized uncertainty principle. 

In contrast to previous studies that focused on the uncertainty trade-offs between conjugated variables, recent works have examined different sources of uncertainty by separating the measured uncertainty into its quantum and classical components~\cite{BiaynickiBirula1975,Luo2003PRL,Luo2005,Korzekwa2014PRA,Yuan2017,Hall2023}.
The quantum component of uncertainty arises when the applied measurement does not commute with the observed quantum system, causing a spread in the measurement outcomes.
This quantum-caused uncertainty can be captured, for instance, by skew information~\cite{Luo2005} and quantum coherence~\cite{Korzekwa2014PRA,Yuan2017} with respect to the observable and the quantum state. 

Spurred by this concept, we propose a novel complementarity relation for a steering witness in this work. 
Specifically, we reveal a complementarity relation between the quantum part (the distillable coherence) and the classical part (the von-Neumann entropy) of a system's total uncertainty. 
We prove that this relation is both tighter and more general compared to the EUR. 
As a direct implication, the quantum-classical uncertainty complementarity relation (QCUR) emerges as a more powerful steering witness, exhibiting superior detection efficiency. 
In addition, from the perspective of coherence distillation tasks, the QCUR violation also quantifies the extra distillable coherence beyond its maximal estimation enabled by steerability, providing a clear operational meaning for the steering witness. 

The proposed steering witness offers two notable advantages. First, it is applicable to systems of \textit{arbitrary} dimensions and \textit{does not} require full-state tomography, significantly enhancing its practicality. 
Second, we demonstrate that the steering witness can \textit{measure} entanglement for all pure bipartite states—a capability that the EUR, the complementarity of coherence for mutually unbiased bases, and Reid's criteria are unable to achieve. This enhanced functionality is further validated through a linear optical experiment presented in this work.

Furthermore, it is known that quantum steering is closely related to measurement incompatibility.
We show that the violation of the QCUR can be used to quantify measurement incompatibility, thereby revealing a deeper connection between the generalized uncertainty principle as well as quantum steering. 
For completeness, we also investigate other properties of the QCUR-based steering witness. 
This includes the asymmetric nature, the ability to detect one-way steering, and its monotonic behavior. 
Our work uncovers a deeper connection between quantum coherence and the uncertainty principle, highlighting its superior utility for steering detection.

% \red{
% \begin{center}
% \begin{tabular}{l l}
%  \hline\hline
%   \multicolumn{1}{c}{Description} & \multicolumn{1}{c}{Expression} \\ [0.1ex] 
%  \hline 
%  Entropic uncertainty relation (EUR)\quad\quad\quad\quad\quad\quad\quad\quad\quad\quad&  $\Hd(\rho) + H_{\Delta'}(\rho) \geq -\log\Omega$\quad\quad\quad\quad\quad\quad\quad\quad\quad\quad\\[0.2ex] 
%  Uncertainty complementarity relation (QCUR) & $C_d(\rho)+S(\rho)=H_\Delta(\rho)$\\[0.2ex] 
%  Steering inequality violation (SIVP) & $\Hd^{B|A}(\mathcal{A}^{\text{LHS}}_{x'}) \geq \Cd_{\text{d}}^{B|A}(\mathcal{A}^{\text{LHS}}_{x}) \quad \forall x,x'$\\[0.2ex] 
%  EUR-based steering inequality & $\Hd(\rho) + H_{\Delta'}(\rho) \geq -\log\Omega$\\[0.2ex] 
%  QCUR-based steering inequality & $H^{B|A}_{\Delta}(\mathcal{A}_x) + H^{B|A}_{\Delta'}(\mathcal{A}_{x'}) \geq -\log \max_{i,j}|\braket{i|j}'|^2$ \\[0.2ex] 
%  \hline\hline
% \end{tabular}
% \end{center}}

\section*{Results}
\subsection*{Distillable coherence and quantum-classical uncertainty complementarity}

In this section, we derive the QCUR from the distillable coherence. To begin with, we provide a concise overview of coherence distillation~\cite{Winter2016PRL}.
Given a priori reference basis $\assemb{\ket{i}}_{i}$, a quantum state $\rho$ is considered incoherent if it is diagonal with respect to the reference basis, i.e., $\rho = \sum_{i} p_{i} \ket{i}\bra{i}$, where $p_i$ forms a probability distribution. 
Thus, states that are not in this form are categorized as coherent states~\cite{Baumgratz2014PRL}. 
We denote the set of incoherent states as $\mathcal{I}$. 
Furthermore, a quantum operation $\Lambda$ is identified as a quantum-incoherent operation (QIO) if it maps an arbitrary incoherent state to another incoherent state. 
For ease of expression, we sometimes extend the term QIOs to refer to the set of quantum incoherent operations.

A coherence distillation process involves the utilization of QIOs to convert $n$ copies of general quantum states into the single-qubit maximally coherent state $\ket{\Phi_{2}}=\sum_{i=0}^{1}\ket{i}/\sqrt{2}$ with a rate $R$.
In the asymptotic limit, i.e., $n \rightarrow \infty$, the maximal rate is called the distillable coherence~\cite{Winter2016PRL}: $\Cd_{\text{d}}(\rho)=\sup \assemb{R:\lim_{n\rightarrow \infty} \inf_{\Lambda \in \text{QIO} } ||\Lambda(\rho^{\bigotimes n})-\Phi_{2}^{\bigotimes Rn}|| =0},$
where $||\bullet||$ denotes the trace norm.
As reported in Ref.~\cite{Winter2016PRL}, the distillable coherence
has a closed form:
\begin{equation}
    \Cd_{\text{d}}(\rho) =  \Hd(\rho) - S(\rho), \label{eq: distillable coherence}
\end{equation}
where $S(\rho)=-\Tr{\rho \log_{2}\rho}$ is the von-Neumann entropy.
Here, we adopt $\Hd(\rho) = S\left[\Delta(\rho)\right]$ as a shorthand notation characterizing the Shannon entropy of the state under the reference basis, where $\Delta(\bullet)=\sum_{i}\ket{i}\bra{i}\bra{i}\bullet\ket{i}$ represents the complete decoherence operation, e.g. $\Delta(\rho) = \sum_i p_i \proj{i}$.
Note that a state $\rho$ is distillable (i.e. $\Cd_{\text{d}} >0$) if and only if $\rho \notin \mathcal{I}$.

To obtain the QCUR, we adopt the notion of ``quantum uncertainty" described in Refs.~\cite{Luo2005,Korzekwa2014PRA,Sun2021PRA}. Specifically, it is known that the von-Neumann entropy $S(\rho)$ characterizes the “classical part of uncertainty" as it aligns with the classical notion, where the uncertainty originates from the lack of information of a system and increases under classical mixing.
In addition, the Shannon entropy $H_\Delta(\rho)$  captures the “total uncertainty" or the observed uncertainty characterized by the probability distribution $\{p_{i}\}_{i}$.
Therefore, according to Eq.~\eqref{eq: distillable coherence}, $C_{\text{d}}(\rho)$ quantifies quantum coherence, i.e. the amount of observed uncertainty that cannot be explained by classical ignorance of the system. 
Along with this reasoning, the distillable coherence can be interpreted as quantum uncertainty.
Through a rearrangement of Eq.~\eqref{eq: distillable coherence}, i.e., $C_{\text{d}}(\rho)+S(\rho)=H_\Delta(\rho)$, one can obtain the QCUR, where the total uncertainty is constituted by the quantum and classical uncertainties~\cite{Luo2005,Girolami2013PRL,Korzekwa2014PRA,Girolami2014PRL,Dressel2014PRA,Angelo2015FP,Sun2021PRA}. 
An equivalent inequality for the QCUR can be expressed as
\begin{equation}
    \Hd(\rho) \geq \Cd_{\text{d}}(\rho), \label{eq: local bound}
\end{equation}
which means that quantum uncertainty $\Cd_{\text{d}}(\rho)$ cannot exceed the total uncertainty $H_{\Delta}(\rho)$. Note that the inequality is saturated when $\rho$ is a pure state, given that there is no classical uncertainty. 
From the perspective of coherence distillation, the QCUR also suggests that the distillable coherence cannot surpass its estimated upper bound quantified by the $\Hd(\rho)$.

\begin{figure}[!htbp]
    \centering
    \includegraphics[width=0.49\columnwidth]{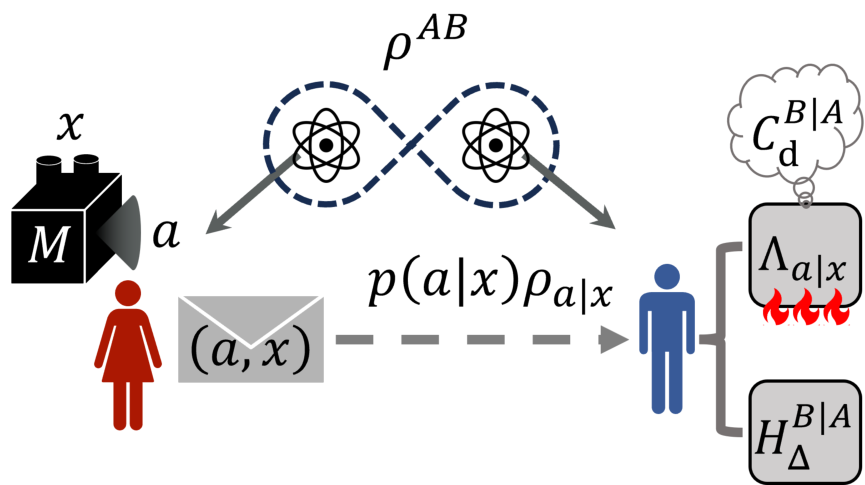}
    \caption{Schematic illustration of the steering-assisted coherence distillation scenario. 
    A bipartite system $\rho^{AB}$ is shared by Alice and Bob.
    Alice measures her subsystem with a measurement setting $x$ and obtains outcome $a$ with probability $p(a|x)$.
    After that, she sends the information $(a,x)$ to Bob through classical communication. 
    Depending on the measurement setting, Bob decides whether to perform coherence distillation by $\Lambda_{a|x}$ or to estimate the maximal distillable coherence by the conditional Shannon entropy $H_{\Delta}^{B|A}$ on the conditional state $\rho_{a|x}$}.
    
    \label{fig:Witnessing steering}
\end{figure}

% \subsection*{Steering-assisted coherence distillation \blue{[JD's suggestion: Violation of quantum-classical uncertainty complementarity by quantum steering]}}
\subsection*{Violation of quantum-classical uncertainty complementarity by quantum steering}

The QCUR holds for all local quantum states in a similar manner to other uncertainty relations. 
As aforementioned, it is known that quantum steering can violate the local uncertainty principle. 
One can therefore expect that steering can also break the QCUR. To formalize this idea, we introduce a steering-assisted coherence distillation task, as described in Fig.~\ref{fig:Witnessing steering}. 
Suppose that Alice and Bob share a bipartite state $\rho^{AB}$. Alice performs a set of positive operator-valued measures (POVM), denoted as $\mathcal{M}=\{M_{a|x}\}_{a,x}$ satisfying $M_{a|x}\geq 0~\forall~a,x$ and $\sum_{a}M_{a|x}=\id~\forall~x$.
Here, $x$ denotes the measurement settings and $a$ represents the corresponding outcomes.
The measurement results can be succinctly represented by a conditional probability distribution $p(a|x)$. After the measurements, Alice communicates both the outcome $a$ and the setting $x$ to Bob, where we denote Bob's conditional state as $\rho_{a|x}$. Conventionally, these results can be summarized by a state assemblage defined by $\mathcal{A} = \{\sigma_{a|x}\}_{a,x}$ with $\sigma_{a|x}=p(a|x)\rho_{a|x}~\forall~a,x$.

It is known that one can employ the local-hidden-state (LHS) model to determine whether a given assemblage is steerable or not. Specifically, an assemblage $\mathcal{A}^{\text{LHS}}$ admits an LHS model when its elements can be described by~\cite{Wiseman2007PRL}:
\begin{equation}
   \sigma^{\text{LHS}}_{a|x} = \sum_{\lambda} p(\lambda)p(a|x,\lambda)\rho_{\lambda} \quad \forall a,x, \label{eq: LHS model} 
\end{equation}
where $\{ \rho_{\lambda} \}_{\lambda}$ and $\{ p(a|x,\lambda) \}_{a,x}$ are, respectively, the hidden states and probabilities that constitute a stochastic process mapping the hidden variable $\lambda$ into the observable outcomes $a|x$.
We note that Alice's choice of $x$ is independent of $\lambda$, referred to as the free-will assumption.
For convenience, we also consider the state assemblage for a fixed setting $x$, denoted as $\mathcal{A}_{x}=\{\sigma_{a|x}\}_{a}$.
Based on the Alice measurement setting, Bob can either distill the quantum coherence or perform projective measurements with the reference bases to obtain the Shannon entropy in order to estimate the maximal distillable coherence. 
Note that with the help of Alice's classical communication, Bob can adjust the local incoherent operation $\Lambda_{a|x}$ to optimize his distillable coherence.
We can then obtain the conditional distillable coherence and the Shannon entropy for a given setting $x$, which are respectively defined as
\begin{equation}
\begin{aligned}
    \Cd_{\text{d}}^{B|A}(\mathcal{A}_{x}) &=  \sum_{a} p(a|x) \Cd_{\text{d}}(\rho_{a|x}), \\
    \Hd^{B|A}(\mathcal{A}_{x}) &= \sum_{a} p(a|x) \Hd \left(\rho_{a|x} \right).\label{eq: Conditional Cd}
\end{aligned}
\end{equation}
By utilizing the convexity of $C_{\text{d}}$, we show that the conditional distillable coherence can be upper-bounded for all LHS models, namely, $\Cd_{\text{d}}^{B|A}(\mathcal{A}^{\text{LHS}}_{x})\leq \sum_{\lambda} p(\lambda) \Hd(\rho_{\lambda})$.
Likewise, the conditional Shannon entropy possesses a lower bound by its concavity, namely $\Hd^{B|A}(\mathcal{A}^{\text{LHS}}_{x})\geq \sum_{\lambda} p(\lambda) \Hd(\rho_{\lambda})$.
We can then derive the QCUR-based steering inequality:
\begin{equation}
      \Hd^{B|A}(\mathcal{A}^{\text{LHS}}_{x'}) \geq \Cd_{\text{d}}^{B|A}(\mathcal{A}^{\text{LHS}}_{x}) \quad \forall~x, x'. \label{eq: steering ineq}
\end{equation}
Consquently, when the assemblage admits LHS model, the condtional distillble coherence (quantum uncertainty) cannot surpass the maximal estimation of distillable coherence (conditional total uncertainty). We remark that the steering inequality generally holds for arbitrary dimensions.

\subsection*{Quantum-classical uncertainty complementarity as a sufficient condition for the EUR}

Upon initial examination, the proposed QCUR and conventional notion of the uncertainty principle appear notably distinct. 
While the former depends on decomposing the uncertainty into classical and quantum components, the latter emphasizes the uncertainty trade-off between incompatible variables or bases. 
Nevertheless, we uncover the interesting connections between these two concepts. 
Specifically, we prove that the QCUR is a sufficient condition for the EUR, which captures the unpredictability of the results from two observables~\cite{Maassen1988,Coles2017Rmp}. 
From the aspect of a steering witness, the QCUR is stronger than the EUR in terms of the detection efficiency. 
In the next section, we further prove that the QCUR-based steering criterion can detect all pure entangled states, while the other criteria cannot.

Recall that the EUR and the EUR-based steering inequality for a pair of non-commuting projective measurements are, respectively, expressed as:
\begin{equation}
    \begin{aligned}
        &\Hd(\rho) + H_{\Delta'}(\rho) \geq -\log\Omega, \\
        \text{and}~~&H^{B|A}_{\Delta}(\mathcal{A}^{\text{LHS}}_x) + H^{B|A}_{\Delta'}(\mathcal{A}^{\text{LHS}}_{x'}) \geq -\log \Omega.
    \end{aligned}
\end{equation}
Here, to keep the notation consistent, we encode the bases for these two measurements performed by Bob ($\{\ket{i}\}_{i}$ and $\{\ket{j}'\}_{j}$) into the pure dephasing maps ($\Delta$ and $\Delta'$) such that $H_{\Delta}(\rho)=\sum_{i}\bra{i}\rho\ket{i} \log \bra{i}\rho\ket{i}$ and $H_{\Delta'}(\rho)=\sum_{j}\bra{j}'\rho\ket{j}' \log \bra{j}'\rho\ket{j}'$. Note that the choice of the bases $\{\ket{i}\}_{i}$ and $\{\ket{j}'\}_{j}$ is independent of Alice’s measurement setting $x$.
In addition, $\Omega = \max_{i,j}|\braket{i|j}'|^2$ denotes the maximal overlap between the two measurement bases. 
The interpretation of this EUR-based steering criterion is that after Alice obtains her measurement data, if she can predict Bob's measurement result with an uncertainty lower than the EUR allows, then Bob's local quantum states, which can reproduce such results, do not exist.

By utilizing the contraction property of the relative entropy and the monotonicity of the logarithm function (see ``Methods" for the detailed derivations), one can show that
\begin{equation}
    \Hd(\rho) \geq \Cd_{\text{d}}(\rho) \geq -H_{\Delta'}(\rho) -\log\Omega.
\end{equation}
Therefore, the QCUR emerges as a sufficient condition for the EUR. Following a similar procedure, one can further deduce that the QCUR-based steering inequality is stronger than the EUR-based steering inequality. Specifically, a state assemblage that violates the QCUR-based steering inequality also violates the EUR-based steering inequality but not \textit{vice versa}.

\subsection*{Properties of the QCUR-based steering inequality violation}
Based on Eq.~\eqref{eq: steering ineq}, quantum steerability enable the conditional distillable coherence to exceed its maximal estimation. Thus, we define the steering inequality violation parameter (SIVP) as  
\begin{equation}
    \mathcal{V}_{\text{S}}(\mathcal{A}):= \max \left\{ \max_x \Cd_{\text{d}}^{B|A}(\mathcal{A}_{x}) - \min_x \Hd^{B|A}(\mathcal{A}_{x}),0 \right\},\label{eq: violation}
\end{equation}
where $\max\{x_{1},x_{2}\}= x_{1},$ if $ x_{1}>x_{2}$; $\max\{x_{1},x_{2}\}= x_{2}$ otherwise.
Accordingly, the SIVP captures the extra conditional distillable coherence beyond its maximal estimation enabled by quantum steering.
Furthermore, we show that the SIVP satisfies the following properties, and the proofs can be found in ``Methods.''

\prop{The SIVP is asymmetric.} 
In the sense that the values of the SIVP are different for Alice to Bob and vice versa.

A steering test should be naturally asymmetrical, and this distinction becomes evident as discussed in previous works~\cite{Bowles2014PRL,Bowles2016PRA,Uola2020RMP}, that permits steering to occur in a unidirectional manner; specifically, from Alice to Bob.

With this property in hand, we can directly show the following.

\prop{The SIVP can detect one-way steering.}

In the steering scenario, Alice and Bob each have distinct roles.
Therefore, the presence of steerability in one direction (e.g., from Alice to Bob) does not guarantee its existence in the opposite direction (from Bob to Alice)~\cite{Bowles2014PRL,Bowles2016PRA}. 
Several examples are provided in ``Methods.''

\prop{The SIVP serves as a full entanglement measure for pure bipartite states.}

It is known that all pure entangled states are steerable~\cite{Wiseman2007PRL,PaulPRL2014}. %This aspect can also be revealed by the SIVP. 
Here, we show that this property can be verified by SIVP. More specifically, for all pure bipartite entangled states $\ket{\psi}^{AB}=\sum_{i} \sqrt{q_i} \ket{i}\ts \ket{i}$, there exists a set of Alice's measurement and a reference basis $\{\ket{i}\}_{i}$ such that $\mathcal{V}_{\text{S}}(\mathcal{A}) = S(\rho_B)$, implying that the SIVP aligns with the entanglement entropy~\cite{Horodecki2009RMP} in this case (see ``Methods” for details).
In other words, the SIVP serves as an entanglement measure for pure bipartite states.
This property underscores the advantage of the SIVP in detecting steerability, as other criteria—such as the EUR, the complementarity of coherence for mutually unbiased bases, and Reid's criteria—fail to achieve this result. 
Moreover, it reveals an operational meaning of the von Neumann entropy: the extra quantum coherence enables by quantum steerability in pure-state scenarios.
Later, we also show the experimental demonstration of the SIVP for pure entangled states.

One can ask whether the SIVP can serve as a steering monotone~\cite{Hsieh2016PRA,Ku2018PRA}. In the most general setting under the resource theory of quantum steering~\cite{Gallego2015PRX}, the answer is negative, because the SIVP does not monotonically decrease under one-way local operations and classical communications (see ``Methods” for details). However, the SIVP could be non-increasing if we restrict the local operations to be QIOs~\cite{Baumgratz2014PRL,Winter2016PRL,Chitambar2016PRL}, as suggested by the numerical results included in ``Methods”. 
Now, we prove that the SIVP is a non-increasing function under genuine incoherent operations (GIOs), which form a subset of the QIOs~\cite{YadinPRX16}.

\prop{The SIVP is a convex-non-increasing function under genuine incoherent operations.}

The main idea of the proof is based on the diagonalization of a GIO w.r.t. the reference basis.
According to Ref.~\cite{YadinPRX16}, there must exist a Kraus representation of the GIO, such that all Kraus operators of the GIO are diagonalized w.r.t. the reference basis. 
Using this property, one can show that the distillable coherence (the Shannon entropy) monotonically decreases (increases) under the GIO, implying that $\mathcal{V}_{\text{S}}$ also monotonically decreases under the GIO. With this property, one can further show that the SIVP is non-increasing under one-way local GIOs and classical communications.

We also emphasize the restriction of GIOs on the local operation in our study makes sense given the fact that the SIVP is based on the distillable coherence. If we allow the most general local operations, the coherence can be distilled by merely changing the local reference basis (see also the discussion in Refs.~\cite{Chitambar2016PRL,Regula2018PRL,Regula2018pra,Fang2018PRL,shiraishi2023alchemy}). In this sense, the general local operation makes a false violation of the SIVP (see also numerical evidence in the ``Methods"). With this property, we can show that the SIVP can be used to quantify measurement incompatibility~\cite{Paul2019PRL,BuscemiPRL2020,Buscemi2023}.
%\red{We note that it is fair to discuss the free operation under the restriction of QIOs or GIOs, given that the SIVP is based on distillable coherence.
%One should ensure the distillable coherence non-increase after the operation; otherwise, the SIVP will contain false violations (We provide some numerical results in the ``Methods'').}

\subsection*{Quantifying measurement incompatibility}
We start by introducing the incompatible measurements.
If multiple physical observables cannot be measured simultaneously, we call these measurements incompatible. 
It is a fundamental characteristic arising from various quantum phenomena, e.g., Bell inequality violation~\cite{Buscemi2012,Brunner2014RMP}, Kochen-Specker contextuality~\cite{Budroni2022RMP}, and uncertainty principles (c.f., section ``Distillable coherence and quantum-classical uncertainty complementarity")~\cite{Busch2014RMP,Baumgratz2014PRL,OtfriedRMP2023}.
Given a set of POVMs $\mathcal{M} = \{M_{a|x}\}_{a,x}$, it is compatible (or jointly measurable) if it can be expressed by 
\begin{equation}
    M_{a|x} = \sum_{\lambda} p(a|x,\lambda)G_{\lambda},\label{eq: joint measurement}
\end{equation}
where $\{G_{\lambda}\}_{\lambda}$ is a parent POVM and $p(a|x,\lambda)$ is a conditional probability.
One can observe that the joint measurable model and the LHS model in Eq.~\eqref{eq: LHS model} are a mathematically similar.
Given a state assemblage, it can be transmitted to a set of POVMs via the concept of the steering-equivalent observables (SEO) $\mathcal{B}=\{B_{a|x}\}_{a,x}$, i.e., $B_{a|x}=(\rho_B)^{-1/2}\sigma_{a|x}(\rho_B)^{-1/2}$, with $\rho_B=\sum_a \sigma_{a|x}$~\cite{Uola2014PRL,Quintino2014PRL,Uola2015PRL}.
We note that once $\rho_B$ was not full rank, the same expression can be obtained by considering an additional isometry (see also Ref.~\cite{Uola2015PRL}). 
One can observe that the SEO is incompatible if and only if the state assemblage is steerable~\cite{Uola2015PRL}.

Inspired by the very recently proposed steering-induced incompatible measure~\cite{Hsieh2023}, we are able to quantify measurement incompatibility by the steering-assisted coherence distillation, namely
\begin{equation}
    \mathcal{V}_{\text{I}}(\mathcal{B}) = \sup_{\rho_B} \mathcal{V}_{\text{S}}[\sqrt{\rho_B}~\mathcal{B} \sqrt{\rho_B}], \label{eq: Steering-induced}
\end{equation}
where $\sup$ is taking over all full-rank states ${\rho_B}$, and $\mathcal{V}_{\text{S}}$ is the SIVP defined in Eq.~\eqref{eq: violation}. We then can show the following:

\prop{}
The optimal steering-assisted coherence distillation $\mathcal{V}_{\text{I}}(\mathcal{B})$ is a valid incompatibility monotone~\cite{Paul2019PRL} in the sense that it satisfies:
(a) $\mathcal{V}_{\text{I}}(\mathcal{B})=0$ if $\mathcal{B}$ is jointly measurable;
(b) $\mathcal{V}_{\text{I}}(\mathcal{B})$ satisfies convexity;
(c) $\mathcal{V}_{\text{I}}(\mathcal{B})$ is non-increasing under post-processing, namely
\begin{equation}
    \{B_{a'|x'}\}_{a',x'}= \mathcal{W}(\{B_{a|x}\}_{a,x})= \sum_{a,x}p(x|x')p(a'|a,x,x')\{B_{a|x}\}_{a,x},\label{eq: wiring map for measurements}
\end{equation}
where $p(x|x')$ and $p(a'|a,x,x')$ are the conditional probabilities and $\mathcal{W}$ is a post-processing scenario defined as a deterministic wiring map~\cite{Gallego2015PRX}.

This result further strengthens the application of the steering-assisted coherence distillation (see “Methods” for details). 
In one direction, it quantitatively connects measurement incompatibility with quantum coherence and gives an additional concrete example for a steering-induced incompatible measure~\cite{Ku2023arxiv}.
In the other direction, we clearly provide a different operational interpretation of measurement incompatibility. Specifically, if we consider $\rho^{AB}$ is a pure entangled state, the SEO $\mathcal{B}$ of $\mathcal{A}$ generated by $\mathcal{M}$ is exactly equivalent to $\mathcal{M}$. 
In this sense, the measurement incompatibility of $\mathcal{M}$ can be accessed in a steering-assisted coherence distillation by properly choosing the pure state $\rho^{AB}$, such that $\mathcal{V}_{\text{I}}(\mathcal{M}) = \sup_{\rho_B} \mathcal{V}_{\text{S}}[\sqrt{\rho_B} \mathcal{M} \sqrt{\rho_B}]$.

\begin{figure*}[!htbp]
\includegraphics[width=0.99\columnwidth]{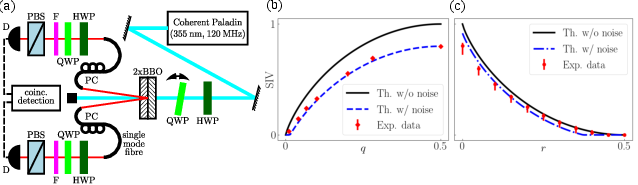}
\caption{\label{fig:setup} (a) Schematic of the experimental setup. 
Individual components are labelled as follows: BBO: $\beta$-BaB$_2$O$_4$ crystal, HWP: half-wave plate, QWP: quarter-wave plate, F: interference bandpass filter (5\,nm spectral width), PBS: polarizer, PC: fiber polarization controller, D: single-photon avalanche photodiode. 
(b) The theoretical predictions are juxtaposed with the experimental results of pure entangled states. 
The black solid curve shows the noise-free theoretical outcomes given by SIVP$(q)=H_{\text{b}}(q)$, which is the binary entropy of the parameter $q$.
Meanwhile, the blue-dashed curve represents the theoretical predictions incorporating the optimal white noise factor $p_{0}=0.026$; the red diamonds indicate experimental results, with error bars obtained via the Monte Carlo method as described in the text. 
(c) The theoretical predictions are compared with the experimental results of the Bell diagonal state.
The black solid curve shows the noise-free theoretical outcomes given by SIVP$(r) = 1 - H_{\text{b}}(r)$; the blue-dashed-dotted curve represents the theoretical predictions incorporating the noise factors $p_{+}=0.009$ and $p_{-}=0.013$ respectively; the red bullets showcase the experimental results.
}
\end{figure*}

\subsection*{Experimental demonstration}
To support the derived theoretical framework, we have performed experimental testing on the platform of linear optics encoding two-qubit states into polarizations of photon pairs. The experimental setup, as depicted in Fig.~\ref{fig:setup}\blue{a}, consists of a laser emitting pulses at 355\,nm that impinge
into a crystal cascade made of two $\beta$-BaB$_2$O$_4$ crystals ($2\times$BBO). These crystals are 1\,mm thick and are mutually positioned, so that their optical axes lie in mutually perpendicular planes \cite{Kwiat_PhysRevA.60.R773}. In these crystals, the laser beam is subjected to the nonlinear process of type-I spontaneous parametric down-conversion. In the first crystal, horizontally polarized laser photons are converted into pairs of vertically polarized photons at 710\,nm. The second crystal facilitates the creation of horizontally-polarized photon pairs from the vertically polarized laser beam. Photon pairs generated in both crystals are subsequently collected into single-mode optical fibers. 
Coherence of the laser beam and indistinguishability in the photons collection assure the effective generation of the photon pairs in a superposition state of both contributing processes, $|\Phi(q,\phi)\rangle =\sqrt{q}|HH\rangle+\mathrm{e}^{i\phi}\sqrt{1-q}|VV\rangle$,
where $H$ and $V$ stand for the horizontal and vertical polarization states, respectively. The parameters $q$ and $\phi$ are controlled by tuning the laser-beam polarization using half- and quarter-wave plates.

The aforementioned single-mode fibers guide the photon pairs to the state detection and the analysis part of the setup. A series of half- and quarter-wave plates followed by a polarizer implement local projections onto any pure polarization state. Such polarization projection is implemented independently on both photons of a pair. We project the photon pairs onto all the combinations of the eigenstates of the Pauli matrices and register the number of simultaneous two-photon detections for all these combinations \cite{Halenkova:12}. A method of maximum likelihood is then used to estimate the most probable density matrix fitting the registered counts \cite{Jezek_PhysRevA.68.012305}. This density matrix is then used to calculate the corresponding SIVP.

To evaluate the experimental uncertainty of the calculated SIVP, we use the fact that registered photon detections follow the Poisson statistics (shot-noise). A Monte-Carlo method is implemented, where all registered counts are randomized assuming the Poisson statistics with the mean value being the actual experimentally observed value. Subsequently the maximum-likelihood method is deployed to estimate the density matrix, which is then used to calculate the SIVP. By repeating this procedure 1000 times, we obtain the statistics of the SIVP under the detection shot-noise and establish the confidence intervals, $\pm\sigma$.

Any experimental implementation is, at least to some degree, imperfect. Partial distinguishability in the generating crystal and imperfections of polarization optics lead to a non-unit purity of the generated states. These imperfections can be reasonably well modeled by white noise. To estimate the amount of such noise, we maximize the expression $F(\rho_{p},\rho_{\text{exp}}) = \max_{p} (\Tr\sqrt{\sqrt{\rho_p}\rho_{\text{exp}}\sqrt{\rho_p}})^2$,
where $F$ denotes the Bures fidelity, $\rho_{\text{exp}}$ is the experimentally observed density matrix and $\rho_p = (1-p) \rho_{\text{th}} + p \id /4$
is the theoretical density matrix $\rho_{\text{th}}$ with added white noise. We have found that for the series of the noisy quasi-pure states $\rho_p$ and presented in Fig.~\ref{fig:setup}\blue{b}, the optimal value of the white-noise factor $p$ is $p_0 = 0.026$ on average.

In Fig.~\ref{fig:setup}\blue{b}, we compare the theoretical predictions with the experimental results.
The noise-free theoretical predictions, i.e., SIVP$(q)=H_{\text{b}}(q)$, where $ H_{\text{b}}(x)=-x\log_{2}x - (1-x)\log_{2}(1-x)$ is the binary entropy, are represented as the black solid curve.
Note that due to the convexity of quantum states, the maximization of Eq.~\eqref{eq: Steering-induced} occurs when Alice and Bob share a pure entangled state.
In our experimental setup, therefore, the measurement incompatibility $\mathcal{V}_{\text{I}}$ can be quantified in terms of steering-assisted coherence distillation when $q=0.5$ [cf., Eq.~\eqref{eq: Steering-induced}].
The predictions with a noise factor $p_{0}=0.026$ are shown with blue-dashed curves.
The experimental results are shown in red diamonds with error bars obtained via the above mentioned Monte Carlo method.

Additionally, we present the results for Bell diagonal states in Fig.~\ref{fig:setup}\blue{c}, obtained through numerical interpolation between two experimentally maximized entangled and mutually orthogonal Bell states: $\ket{\Phi^{+}}=\ket{\Phi(1/2,0)}$ and $\ket{\Phi^{-}}=\ket{\Phi(1/2,\pi)}$. A Bell diagonal state is defined as $\rho_{\text{Bell}}=r\ket{\Phi^{+}}\bra{\Phi^{+}} + (1-r)\ket{\Phi^{-}}\bra{\Phi^{-}}$. The noise-free theoretical predictions are expressed as SIVP$(r) = 1 - H_{\text{b}}(r)$, depicted by the black solid curve. 
The experimental setup was meticulously calibrated to generate the purest states possible, reducing noise to $p_{+} = 0.009$ and $p_{-} = 0.013$, represented by the blue-dashed-dotted curve. 
Finally, our experimental results for the Bell diagonal states are shown as red bullets with error bars.

\section*{Discussion}\label{sec_discussion}
In this work, we proposed a novel complementarity relation, termed the QCUR for steering detection. 
We prove that the QCUR is a stronger condition than the EUR for detecting steering. 
Further, the QCUR emerges as the most efficient steering witness compared with the EUR, the complementarity of coherence for mutually unbiased bases, and Reid's criteria. 
Notably, we show that the QCUR-based steering witness is capable of measuring entanglement for pure bipartite states. 
This theoretical result is further validated through a linear optical experiment, underscoring the practical applicability of the QCUR in experimental setups.
From the operational aspects, the QCUR-based steering inequality can be implemented in coherence distillation tasks \textit{without requiring full state tomography}, and its violation quantifies the extra conditional distillable coherence beyond its maximal estimation enabled by quantum steering.

Furthermore, we showed that violation of the QCUR can also be used to quantify measurement incompatibility. 
It has several mathematical properties, including asymmetry, capable of detecting one-way steering, and is monotonic under GIOs.

For future directions, several open questions arise naturally from our work. For instance, can the proposed distillation scenarios also detect and potentially quantify quantum steering, and then extended to multi-party steering? Another intriguing avenue is whether our framework can be adapted to the one-shot~\cite{Regula2018PRL,Qi2019IEEE} and asymptotic reversibility settings~\cite{Eric2018PRA}.
Moreover, Ref.~\cite{Berta2010} demonstrates that entanglement with a quantum memory can enable more precise predictions of measurement outcomes, effectively violating the usual uncertainty relation. It would be interesting to explore how this approach can be generalized from device-dependent to one-sided device-independent scenarios.

\section*{Methods}\label{sec_method}
\subsection*{Derivation of the QCUR-based steering inequality}\label{apx: Proof of the steering inequality}
For convenience, we define the optimal conditional distillable coherence and the conditional Shannon entropy as
\begin{equation}
    \CCd_{\text{d}}(\mathcal{A}) := \max_x \Cd_{\text{d}}^{B|A}(\mathcal{A}_{x})\quad \text{and} \quad \CH(\mathcal{A}) :=\min_x \Hd^{B|A}(\mathcal{A}_{x}), \label{eq: optimal siv}
\end{equation}
respectively.

\begin{proof}---The inequality $\CCd_{\text{d}}(\mathcal{A}) \leq \CH(\mathcal{A})$ holds if the assemblage $\mathcal{A}$ admits an LHS model.
    
Considering that the assemblage received by Bob can be described by the LHS model, the upper bound of the coherence distillation can be readily obtained:
\begin{equation}
\begin{aligned}
    \CCd_{\text{d}}(\mathcal{A}) 
    &= \max_{x} \sum_{a} p(a|x) \Cd_{\text{d}} \left[ \sum_{\lambda}p(\lambda|a,x)\rho_{\lambda} \right] \\
    &= \max_{x} \sum_{a} p(a|x) \Cd_{\text{d}} \left[ \sum_{\lambda}\frac{p(a|x,\lambda)p(\lambda)}{p(a|x)}\rho_{\lambda} \right] \\
    &\leq \max_{x} \sum_{a} p(a|x) \sum_{\lambda}\frac{p(a|x,\lambda)p(\lambda)}{p(a|x)} \Cd_{\text{d}}(\rho_{\lambda}) \\
    &= \max_{x} \sum_{\lambda} p(\lambda) \Cd_{\text{d}}(\rho_{\lambda}) \\
    &= \sum_{\lambda} p(\lambda) \Cd_{\text{d}}(\rho_{\lambda}) \\
    &\leq \sum_{\lambda} p(\lambda) \Hd(\rho_{\lambda}).
\end{aligned}
\end{equation}
Since $\Hd(\rho)$ is concave in $\rho$, we readily obtain $\sum_{\lambda} p(\lambda) \Hd(\rho_{\lambda}) \leq \CH(\mathcal{A})$ by an analogous derivation.
This concluds the proof of steering inequality:
\begin{equation}
     \CCd_{\text{d}}(\mathcal{A}) \leq \CH(\mathcal{A})\quad \text{if}~ \mathcal{A}\in \text{LHS}.
\end{equation}
\end{proof}

\subsection*{Proof for the QCUR as a sufficient condition for the EUR}
We first compare the QCUR and EUR, showing that one can relax the lower bound of the QCUR to obtain the EUR, that is
\begin{equation}
\begin{aligned}
    \Hd(\rho) &\geq \Cd_{\text{d}}(\rho)  \\
    &= D\left[\rho||\Delta(\rho)\right] \\ 
    &\geq D\left[\Delta'(\rho)||\Delta' \circ \Delta(\rho)\right] \\ 
    &= \Tr~\Delta'(\rho) \left[\log \Delta'(\rho)-\log \Delta' \circ \Delta(\rho)\right] \\ 
    &= -H_{\Delta'}(\rho) - \sum_{j}\bra{j}'\rho\ket{j}'\log \sum_{i} |\braket{i|j}'|^2 \bra{i}\rho_{a|x'}\ket{i} \\ 
    &\geq -H_{\Delta'}(\rho) - \sum_{j}\bra{j}'\rho\ket{j}'\log~ \left(\max_{i,j}|\braket{i|j}'|^2 \right) \sum_{i}\bra{i}\rho_{a|x'}\ket{i} \\ 
    &= -H_{\Delta'}(\rho) - \log \Omega,
\end{aligned}
\end{equation}
where $D(\rho||\sigma) = \Tr~\rho\left( \log \rho - \log \sigma \right)$ is the relative entropy, $\Delta'(\rho) = \sum_{j} \bra{j}'\rho\ket{j}' \ket{j}'\bra{j}'$ is an arbitrary complete decoherence operation in the basis $\{\ket{j}'\}_{j}$, and $\Omega = \max_{i,j}|\braket{i|j}'|^2$ represents the maximal overlap between two different reference bases.
Note that the second inequality is given by the contraction of the relative entropy~\cite{Coles2017Rmp} and the third inequality is due to the applied maximization.
The result indicates that satisfying the QCUR is a sufficient condition of the EUR.
% \begin{equation}
% \begin{aligned}
%     \Cd_{\text{d}}(\rho) &= H_{\Delta}(\rho) - S(\rho) \\ 
%     &= -S(\rho) + S\left[\Delta\left( \sum_{j} E_{j} \ket{E_{j}}\bra{E_{j}} \right) \right] \\ 
%     &= -S(\rho) + S\left[\sum_{i} \left( \sum_{j} E_{j}|\braket{i|E_{j}}|^2 \right)  \ket{i}\bra{i}\right] \\ 
%     &= -S(\rho) - \sum_{i}  \left( \sum_{j} E_{j}|\braket{i|E_{j}}|^2 \right) \log  \left( \sum_{k} E_{k}|\braket{i|E_{k}}|^2 \right) \\
%     &\geq -S(\rho) - \sum_{i}  \left( \sum_{j} E_{j}|\braket{i|E_{j}}|^2 \right) \log \max_{i,j}|\braket{i|E_{j}}'|^2 \\
%     &= -S(\rho) -\log \max_{i,j}|\braket{i|E_{j}}'|^2
% \end{aligned}
% \end{equation}

Next, we show that the QCUR-based steering inequality is stronger than the EUR-based steering inequality~\cite{Coles2017Rmp}.
We start from the steering inequality in Eq.~\eqref{eq: steering ineq} by taking two Alice's measurement settings $x$ and $x'$, we have
\begin{equation}
\begin{aligned}
    \sum_{a}p(a|x)\Hd(\rho_{a|x}) &\geq \sum_{a}p(a|x')\Cd_{\text{d}}(\rho_{a|x'}) \quad \forall~x,x'\\
    &= \sum_{a}p(a|x')\left\{ D\left[\rho_{a|x'}||\Delta (\rho_{a|x'}) \right]\right\} \\
    &\geq  \sum_{a}p(a|x')  \left\{D\left[\Delta' (\rho_{a|x'})||\Delta' \circ \Delta (\rho_{a|x'})\right]\right\} \\
    &= \sum_{a}p(a|x') \Tr~ \Delta' (\rho_{a|x'}) \left[ \log \Delta' (\rho_{a|x'}) - \log \Delta' \circ \Delta (\rho_{a|x'}) \right] \\
    &= -\sum_{a}p(a|x') H_{\Delta'}(\rho_{a|x'}) - \sum_{a} p(a|x')\sum_{j}\bra{j}'\rho_{a|x'}\ket{j}'\log \sum_{i} |\braket{i|j}'|^2 \bra{i}\rho_{a|x'}\ket{i} \\
    &\geq -\sum_{a}p(a|x') H_{\Delta'}(\rho_{a|x'}) - \log\Omega.
     \label{eq: ent-based steer eq}
\end{aligned}
\end{equation}
The above result directly implies the EUR for unsteerable states~\cite{Coles2017Rmp}, i.e.,
\begin{equation}
    \sum_{a} p(a|x) H_{\Delta}(\rho_{a|x}) + \sum_{a} p(a|x') H_{\Delta'}(\rho_{a|x'}) \geq - \log\Omega. \label{eq: EUR}
\end{equation}
% which states that if Alice can predict Bob's measurement results with the uncertainty lower than $- \log\Omega$, then Bob cannot reproduce these results by his local quantum states, reads
% \begin{equation}
%     H_{\Delta}(\rho_B) + H_{\Delta'}(\rho_B) \geq \sum_{a} p(a|x) H_{\Delta}(\rho_{a|x}) + \sum_{a} p(a|x') H_{\Delta'}(\rho_{a|x'}).
% \end{equation}
Therefore, we conclude that the QCUR-based inequality in Eq.~\eqref{eq: steering ineq} is a stronger criterion compared to the EUR-based one.

For completeness, we provide a detail derivation from the conventional conditional Shannon entropy $H(\mathbb{B}|\mathbb{A})$, applying the notation used in this paper, i.e., $H_{\Delta}^{B|A}$.
Consider the product measurements $\mathbb{A}\ts \mathbb{B}$ for two projective measurements $\mathbb{A}=\{\ket{\mathbb{A}_{a}}\bra{\mathbb{A}_{a}}\}_{a}$ and $\mathbb{B}=\{\ket{\mathbb{B}_{b}}\bra{\mathbb{B}_{b}}\}_{b}$ performed by Alice and Bob on their own systems, the joint distribution $p(a,b)$ and the conditional Shannon entropy $H(\mathbb{B}|\mathbb{A})$ read:
\begin{equation}
\begin{aligned}
    H(\mathbb{B}|\mathbb{A}) &= H(\mathbb{A},\mathbb{B})-H(\mathbb{A}) \\
    &= \sum_{a,b}p(a,b|\mathbb{A},\mathbb{B})\log p(a,b|\mathbb{A},\mathbb{B}) - \sum_{a}p(a|\mathbb{A})\log p(a|\mathbb{A}) \\
    &= \sum_{a,b} p(a|\mathbb{A})p(b|a,\mathbb{A},\mathbb{B})[\log p(b|a,\mathbb{A},\mathbb{B}) + \log p(a|\mathbb{A}) ] - \sum_{a}p(a|\mathbb{A})\log p(a|\mathbb{A}) \\
    &= \sum_{a,b} p(a|\mathbb{A})p(b|a,\mathbb{A},\mathbb{B})\log p(b|a,\mathbb{A},\mathbb{B}) \\
    &= \sum_{a} p(a|\mathbb{A}) \sum_{b}\bra{\mathbb{B}_{b}}\rho_{a|\mathbb{A}}\ket{\mathbb{B}_{b}} \log \bra{\mathbb{B}_{b}}\rho_{a|\mathbb{A}}\ket{\mathbb{B}_{b}},
\end{aligned}
\end{equation}
where we can find that $H(\mathbb{B}|\mathbb{A})$ is equivalent to $H_{\Delta}^{B|A}$ by choosing $\Delta(\rho)=\sum_{b}\bra{\mathbb{B}_{b}}\rho\ket{\mathbb{B}_{b}}$.

\subsection*{Proof of Property 1: Asymmetry of the steering inequality violation}\label{apx: Proof of property 1}
\begin{proof}---To demonstrate the asymmetry, we consider general two-qudit states defined by
\begin{equation}
\chi(\Vec{r},\Vec{s},\Vec{t}) = \frac{1}{4}\left( \id \ts \id + \Vec{r} \cdot \Vec{\sigma} \ts \id + \id \ts \Vec{s} \cdot \Vec{\sigma} + \sum_{i,j=1}^{3} t_{ij} \sigma_{i} \ts \sigma_{j} \right), \label{eq: two-qubit system}
\end{equation}
where $\{\Vec{r},\Vec{s},\Vec{t}\}$ are the vectors with norm less than unit and $\Vec{\sigma} = (\sigma_{1},\sigma_{2},\sigma_{3})$ is the Pauli vector.
The SIVP is asymmetric, if the value of the SIVP depends on the local Bloch vector, i.e., $\Vec{r}$ and $\Vec{s}$.
Here, we consider Alice performing measurements described by $M_{a|x} = [\id + (-1)^{a} \sigma_{x} ]/2$, with $a \in \{0,1 \}$ and $x \in \{1,3 \}$; where $\sigma_1$ and $\sigma_3$ are the Pauli X and Z matrices, respectively, then the assemblage received by Bob becomes
\begin{equation}
\begin{aligned}
    \sigma_{a|x}
    &= \TrA \left[  M_{a|x} \ts \id \chi(\Vec{r},\Vec{s},\Vec{t})  \right]\\
    &=  \frac{1}{4} \TrA\left[ M_{a|x}\ts\id + M_{a|x}\Vec{r}\cdot \Vec{\sigma}\ts \id +  M_{a|x}\ts \Vec{s}\cdot \Vec{\sigma} +\sum_{i,j=1}^{3} M_{a|x}t_{ij} \sigma_{i} \ts \sigma_{j}\right]\\
    &= \frac{1}{4} \left\{ \id + \frac{1}{2} \Tr \left[ \Vec{r}\cdot \Vec{\sigma} + (-1)^{a}\sigma_{x} \Vec{r}\cdot \Vec{\sigma} \right] \id + \Vec{s}\cdot \Vec{\sigma} + \frac{1}{2} \sum_{i,j=1}^{3}t_{ij} \Tr\left[ \sigma_{i} + (-1)^{a} \sigma_{x}\sigma_{i} \right] \sigma_{j} \right\}\\
    &= \frac{1}{4}\left[ \id + (-1)^{a}r_{x}\id + \Vec{s}\cdot \Vec{\sigma} + \sum_{j}(-1)^{a} t_{xj}\sigma_{j} \right] \\
    &= \frac{1}{2}\left[ 1 + (-1)^{a}r_{x}\right] \times \frac{1}{2}\left[ \id +\sum_{j} \frac{  s_{j} + (-1)^a t_{xj}  }{1+(-1)^{a}r_{x}}\sigma_{j} \right], \label{eq: two-qubit assemb}
\end{aligned}
\end{equation}
with probabilities $\Tr\sigma_{a|x} = [ 1 + (-1)^{a}r_{x}]/2$.
To obtain the SIVP, we need to calculate the eigenvalue of the reduced state $\rho_{a|x}$ and its dephased counterpart $\Delta(\rho_{a|x})=\sum_{i=0}^{1}\bra{i}\rho_{a|x}\ket{i}\ket{i}\bra{i}$, which are, respectively, 
\begin{equation}
    E_{a|x,\pm}(r,s,t) = \frac{1}{2} \left[ 1 \pm \frac{ \sqrt{\sum_{j} \left[ s_{j} + (-1)^a t_{xj}\right]^{2} } }{ 1 + (-1)^{a}r_{x} } \right] \quad \text{and} \quad
    E^{\text{deph}}_{a|x,\pm}(r,s,t) = \frac{1}{2}\left[ 1 \pm \frac{s_{3}+(-1)^{a}t_{x}\delta_{x,3}}{1+(-1)^{a}r_{x}}  \right].
\end{equation}
Here, we can observe that the SIVP in Eq.~\eqref{eq: optimal siv} depends on the local Bloch vectors $\Vec{r}$ and $\Vec{s}$.
As we swap the general two-qubit states $\text{SWAP}[\chi(\Vec{r},\Vec{s},\Vec{t})] = \chi(\Vec{s},\Vec{r},\Vec{t})$, we obtain different values of SIVP, which states that the SIVP is asymmetric.
\end{proof}

Note that the proof can be generalized to an arbitrary $d$-dimensional system by utilizing the generalized Pauli matrices $\Vec{\sigma} = (\sigma_{1},\sigma_{2},...,\sigma_{d^2-1})$.
By the same method in Eq.~\eqref{eq: two-qubit assemb}, one finds that the assemblage depends on $\vec{r}$ and $\vec{s}$, expressed as:
\begin{equation}
    \sigma_{a|x} = \frac{1}{d^2} \left\{  \left(1+ \Tr M_{a|x} \vec{r}\cdot \vec{\sigma} \right)\mathbb{1} + \sum_{j,k=1}^{d^2 -1}  \left[ s_{k} + \Tr (M_{a|x} \sigma_{j}) t_{jk} \right]\sigma_{k} \right\},
\end{equation}
with probabilities $p(a|x)=(1+ \Tr M_{a|x} \vec{r}\cdot \vec{\sigma})/d$ only depends on $\vec{r}$.
Although computing the eigenvalues in a $d$-dimensional system is a non-trivial task, it can be shown that the eigenvalues of the reduced state and its dephased counterpart still depend on the vectors $\vec{r}$ and $\vec{s}$. 
Consequently, the SIVP, which is derived from these conditional probabilities and eigenvalues, changes when the parameters of $\vec{r}$ and $\vec{s}$ are interchanged, indicating that the SIVP is asymmetric.

\subsection*{Example of Property 2: Detecting one-way steering}
\begin{figure}[!htbp]
    \centering
    \includegraphics[width=0.45\columnwidth]{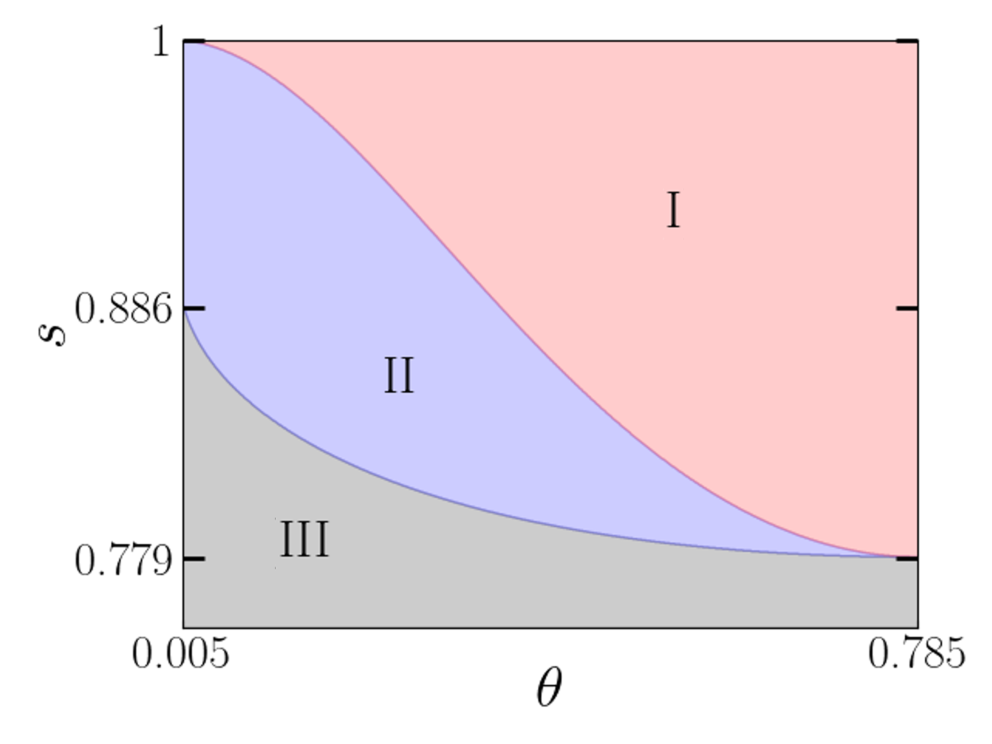}
    \caption{Ability of the steering inequality violation to demonstrate one-way quantum steering in the parameter windows: $s\in[0.75, 1]$ and $\theta \in [0.005, 0.25\pi]$ for the states in Eq.~\eqref{eq: 2qbitstates}.
    The light-red area (I) represents both $\mathcal{V}_{\text{S}}(\mathcal{A}^{B \rightarrow A})>0$ and $\mathcal{V}_{\text{S}}(\mathcal{A}^{A \rightarrow B})>0$, suggesting that quantum steering occurs in both directions.
    The light-blue area (II), conversely, represents $\mathcal{V}_{\text{S}}(\mathcal{A}^{B \rightarrow A})>0$ and $\mathcal{V}_{\text{S}}(\mathcal{A}^{A \rightarrow B})=0$, indicating that the respective steerable state only allows Bob to steer Alice.
    Finally, the grey area (III) represents the range of the parameters $s,q$ such that $\mathcal{V}_{\text{S}}(\mathcal{A}^{B \rightarrow A})=0$ and $\mathcal{V}_{\text{S}}(\mathcal{A}^{A \rightarrow B})=0$.
    }\label{fig: Asymmetry}
\end{figure}

As a concrete example, we now present the SIVP of a set of states described by 
\begin{equation}
    % \chi(s,q) = s \ket{\psi_q}\bra{\psi_q} + (1-s) \TrB \left(\ket{\psi_q}\bra{\psi_q}\right)\ts \id/2, \label{eq: 2qbitstates}
    \chi(s,\Vec{q}) = s\ket{\psi_{\vec{q}}}\bra{\psi_{\vec{q}}}
    + (1-s) \text{Tr}_{B} [\ket{\psi_{\vec{q}}}\bra{\psi_{\vec{q}}}] \otimes \frac{\mathbb{1}}{d}, \label{eq: 2qbitstates}
\end{equation} 
where $\ket{\psi_{\vec{q}}}=\sum_{i}^{d=2} \sqrt{q_{i}}\ket{i}\ket{i}$ and $\vec{q}=(\text{cos}^{2}\theta,\text{sin}^{2}\theta)$
% $\vec{q}=(q_1,q_2,...,q_d)$.
% To demonstrate the one-way steering detection, we consider $\vec{q}=(q,1-q)$
% $\ket{\psi_q} =\sum_{i=0}^{1} t_i(q) \ket{i}\ts \ket{i}$, $t_{0}(q) = q$, and $t_{1}(q) = 1-q$ 
in the parameter windows: $s\in [0.75,1]$ and $\theta \in [0.005,0.25\pi]$.
The SIVP values are shown in Fig.~\ref{fig: Asymmetry}.
In the light-red area (I), the steerability can be detected from both directions, i.e., $\mathcal{V}_{\text{S}}(\mathcal{A}^{B \rightarrow A})>0$ and $\mathcal{V}_{\text{S}}(\mathcal{A}^{A \rightarrow B})>0$. However, in the light-blue area (II), one finds $\mathcal{V}_{\text{S}}(\mathcal{A}^{B \rightarrow A})=0$, while $\mathcal{V}_{\text{S}}(\mathcal{A}^{B \rightarrow A})>0$, which indicates that the SIVP is only witnessed from Bob to Alice. 
Finally, in the grey area (III), the steerability cannot be detected from any direction.
We note that $\chi(s,\Vec{q})$ can be generalized to higher dimensions by considering $\vec{q}=(q_1,q_2,...,q_d)$.

\subsection*{Proof of Property 3: Detecting all pure entangled states}\label{apx: Proof of property 3}

In the following, we show that when the bipartite state is pure, the SIVP is a full entanglement measure for pure states.
% We separate the derivation into two parts, which are given in the following:

\begin{proof}---
Given that Alice and Bob share a bipartite pure state, we assume that Alice performs rank-1 POVMs, ensuring that the resulting reduced states $\rho_{a|x}$ are pure.
% First, we demonstrate that the best measurements Alice can perform are rank-1 POVMs.
% We consider a fine-grained POVM $\{M^{i}_{a|x}\}$ s.t. $\sum_{i} M^{i}_{a|x}=M_{a|x}~\forall a,x$, where each element satisfies $M^{i}_{a|x}\geq 0$ and $\text{rank}(M^{i}_{a|x})=1$, performed by Alice.
% Under this new POVM, Bob’s new assemblage is obtained by averaging over the index $i$, i.e., $\sum_{i} p(a,i|x)\rho_{a,i|x} = p(a|x)\rho_{a|x}$.
% This provides a better rate of distillable coherence on Bob's side, reads
% \begin{equation} 
% \begin{aligned} 
%     C^{B|A}_{\text{d,fine}}(\mathcal{A}_{x}) 
%     &= \sum_{i,a} p(a,i|x) C_{\text{d}}(\rho_{a,i|x}) \\
%     &= \sum_{a} p(a|x)\sum_{i} \frac{p(a,i|x)}{p(a|x)} C_{\text{d}}(\rho_{a,i|x}) \\
%     &\geq \sum_{a} p(a|x) C_{\text{d}}\left[ \sum_{i} \frac{p(a,i|x)}{p(a|x)}\rho_{a|x}\right] \\
%     &= \sum_{a} p(a|x) C_{\text{d}}(\rho_{a|x}) \\
%     &=C^{B|A}_{\text{d}}(\mathcal{A}_{x}), 
% \end{aligned} 
% \end{equation} where the inequality follows from convexity.
% Similarly, one can easily show that $H_{\Delta,\text{fine}}^{B|A}(\mathcal{A}_{x}) \leq H_{\Delta}^{B|A}(\mathcal{A}_{x})$ due to concavity.
% This result suggests that to obtain the largest violation, the optimal measurements performed by Alice can be considered as rank-1 POVMs.
Under this consideration, the reduced states $\rho_{a|x}$ from any pure bipartite state are pure. 
This implies that the conditional distillable coherence can be reached by the pure-state decomposition $\{p(a|x), \ket{\psi_{a|x}}\}_{a,x}$, such that $\sum_{a}p(a|x) \ket{\psi_{a|x}}\bra{\psi_{a|x}}=\rho_{B}~\forall x$, i.e.,
\begin{equation}
\begin{aligned}
    C_{\text{d}}^{\star}(\mathcal{A}) 
    &= \max_{ \{ p(a|x),\ket{\psi_{a|x}} \} } \sum_{a} p(a|x) C_{\text{d}}(\ket{\psi_{a|x}}) \\
    &= \max_{ \{ p(a|x),\ket{\psi_{a|x}} \} } \sum_{a} p(a|x) H_{\Delta}(\ket{\psi_{a|x}}) \\
    &= H_{\Delta}\left(\sum_{a} p(a|x^{\uparrow})\ket{\psi_{a|x^{\uparrow}}}\bra{\psi_{a|x^{\uparrow}}}\right) \\
    &= H_{\Delta}(\rho_{B}),
\end{aligned}
\end{equation}
where the third line is given by the concave roof and the optimal $x=x^{\uparrow}$.
Similarly, we have
\begin{equation}
\begin{aligned}
    H_{\Delta}^{\star}(\mathcal{A}) 
    &= \min_{ \{ p(a|x),\ket{\psi_{a|x}} \} } \sum_{a} p(a|x) H_{\Delta}(\ket{\psi_{a|x}}) \\
    &= \min_{ \{ p(a|x),\ket{\psi_{a|x}} \} } \sum_{a} p(a|x) \left[ H_{\Delta}(\ket{\psi_{a|x}}) - S(\ket{\psi_{a|x}}) \right] \\
    &= C_{\text{d}}\left(\sum_{a} p(a|x^{\downarrow})\ket{\psi_{a|x^{\downarrow}}}\bra{\psi_{a|x^{\downarrow}}}\right) \\
    &= C_{\text{d}}(\rho_{B}),
\end{aligned}
\end{equation}
where the third line is obtained by the convex roof and the optimal $x=x^{\downarrow}$.
By the definition of the SIVP, we obtain
\begin{equation}
    \mathcal{V}_{\text{S}}(\mathcal{A}):= \max\left\{C_{\text{d}}^{\star}(\mathcal{A}) - H_{\Delta}^{\star}(\mathcal{A}) , 0\right\} = S(\rho_{B}).
\end{equation}
We note that $S(\rho_{B})$ is the entanglement entropy, which is a valid entanglement measure for pure states~\cite{Horodecki2009RMP}, and thus we conclude the proof.
\end{proof}

\subsection*{Monotonicity under genuine incoherent operations}\label{apx: Proof of property 5}

Quantum steering has been articulated within the resource theory framework~\cite{Gallego2015PRX}. A measure $\mathcal{S}$ qualifies as a convex steering monotone if it adheres to the following properties:
\begin{itemize}
    \item[(i)] $\mathcal{S}(\sigma_{a|x})=0$ for all $\sigma_{a|x}\in~\text{LHS}$.
    \item[(ii)] $\mathcal{S}\left[p\sigma_{a|x}+(1-p)\sigma^{'}_{a|x}\right]\leq p\mathcal{S}{\sigma_{a|x}} + (1-p)\mathcal{S}(\sigma^{'}_{a|x})$, for any real number $0\leq p\leq 1$ and assemblages $\sigma_{a|x}$ and $\sigma^{'}_{a|x}$.
    \item[(iii)] Non-increasing under one-way local operations and classical communication:
    \begin{equation}
        \sum_{\xi} p(\xi)\mathcal{S}\left[ \frac{\Xi_{\xi} (\sigma_{a|x})}{ \Tr \Xi_{\xi}(\sigma_{a|x}) } \right] \leq \mathcal{S}(\sigma_{a|x})\quad \forall \sigma_{a|x}, 
    \end{equation}
    where $p(\xi)=\Tr \Xi_{\xi}(\sigma_{a|x})$ and $\sum_{\xi}p(\xi)=1$.
\end{itemize}
It is clear that $\mathcal{V}_{\text{S}}$ satisfies properties (i) and (ii).
Nonetheless, property (iii) is satisfied only under limited local operations.
Property (iii) states that quantum steering should not increase under free operations, e.g., one-way local operations and classical communication~\cite{Gallego2015PRX}.

In the scenario of steering-assisted coherence distillation, local operations must also adhere to incoherent operations. 
In the following, we consider that these local operations belong to the set of genuine incoherent operations~\cite{YadinPRX16}, which reside as a subset within incoherent operations~\cite{Chitambar2016PRL}.

\begin{figure}[!htbp]
    \centering
    \includegraphics[width=0.5\columnwidth]{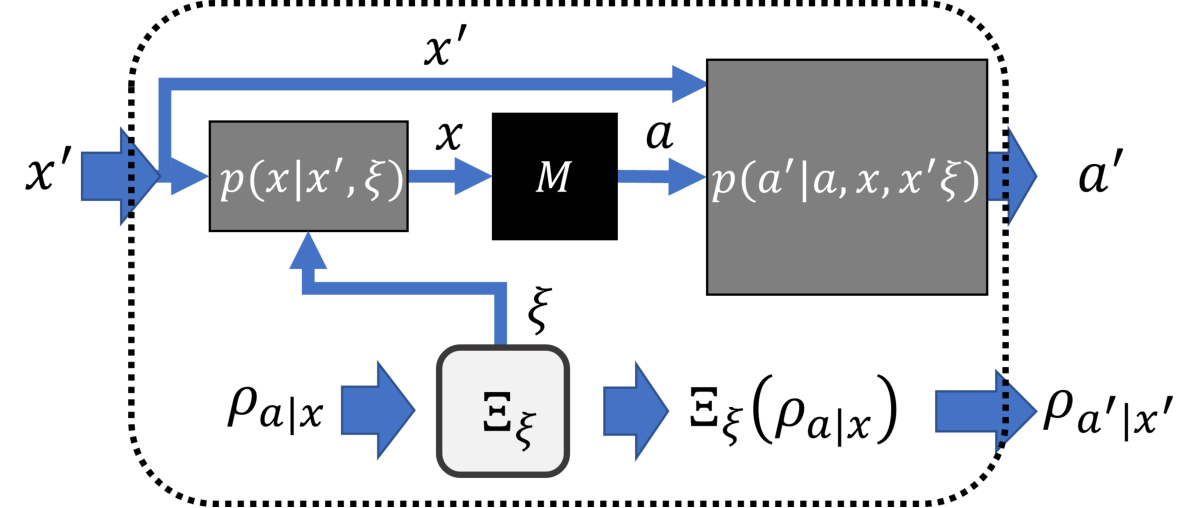}
    \caption{Schemetic illustration of one-way local genuinely incoherent operations and classical communication.
    The final assemblage $\sigma_{a'|x'} = \Xi(\sigma_{a'|x'})$ is given by a black box (represented by black-dashed rectangle) with inputs $x'$ and outputs $a'$.
    To do this, Bob first applied a genuinely incoherent operation $\Xi_{\xi}$ with probability $p(\xi)$ on his subsystem $\rho_{a|x}$ and obtain the state $\rho_{a'|x'} = \Xi_{\xi} (\rho_{a|x})$. 
    Thereafter, he send $\xi$ to Alice through classical communication.
    Based on the information of $x'$ and $\xi$, Alice performs a local wiring map (the first gray box) with probability distribution $p(x|x',\xi)$ generating inputs $x$ for her measurements $M$. 
    Then, Alice receives outcomes $a$ and processes all classical labels $a$, $x$, $x'$, and $\xi$ to generate the final outputs $a'$ by a local wiring map (the second gray box) with a probability distribution $p(a'|a,x,x',\xi)$.
    }
    \label{fig: LQICC Map}
\end{figure}

As shown in Fig.~\ref{fig: LQICC Map}, a local stochastic genuine incoherent operation is performed on Bob's system. 
Specifically, Bob introduces a device that generates a random outcome $\xi$ with probability $p(\xi)$. After receiving the outcome, Bob sends his system into a corresponding genuinely incoherent operation $\Xi_\xi$~\cite{YadinPRX16}, which can be characterized by the set of Kraus operators $\{K_{k,\xi}\}_{k}$, such that each Kraus operator is diagonalized under the reference
basis.
Additionally, the outcome $\xi$ is also sent to Alice through classical communication, so that she utilizes classical stochastic maps defined by $\{p(a'|a,x,x',\xi), p(x|x',\xi)\}$ to post-process her measurement results. Consequently, the entire process $\Xi$ transforms the initial assemblage $\sigma_{a|x}$ into a final assemblage $\sigma_{a'|x'}$:
\begin{equation}
    \sigma_{a'|x'} 
    = \Xi(\sigma_{a|x})
    = \sum_{a,x,\xi}p(x|x',\xi)p(a'|a,x,x',\xi) p(\xi)\Xi_\xi (\sigma_{a|x}). \label{eq: Big Xi}
\end{equation}

Now, we prove that the SIVP is monotonic under this constrain.
% The steering inequality violation is non-increasing under one-way local genuinely incoherent operations and classical communication.
\begin{proof}---We aim to use the strategy in Eqs.~\eqref{proof: post-process monotone Cd} and \eqref{proof: post-process monotone CH} by first showing that $\CCd_{\text{d}}$ is non-increasing after the process, namely
\begin{equation}
\begin{aligned}
    \CCd_{\text{d}}\left[\Xi(\mathcal{A})\right]
    &= \max_{x'} \sum_{a'} p(a'|x') \Cd_{\text{d}} \left[\frac{\sigma_{a'|x'}}{p(a'|x')}\right] \\
    &= \max_{x'} \sum_{a'} p(a'|x') \Cd_{\text{d}}\left[\sum_{a,x,\xi} \frac{  p(x|x',\xi)p(a'|a,x,x',\xi) p(\xi)K_{\xi}\sigma_{a|x}K_{\xi}^{\dag} }{p(a'|x')}\right] \\
    &= \max_{x'} \sum_{a'} p(a'|x') \Cd_{\text{d}}\left[\sum_{a,x,\xi} \frac{ p(x|x',\xi)}{p(a'|x')} \frac{p(a'|x',\xi)p(x|a',x',\xi)p(a|x,a',x',\xi)}{p(x|x',\xi)p(a|x,x',\xi)}p(a|x)p(\xi)K_{\xi}\rho_{a|x}K_{\xi}^{\dag}\right] \\
    &= \max_{x'} \sum_{a'} p(a'|x') \Cd_{\text{d}}\left[\sum_{a,x,\xi} \frac{p(a'|x',\xi)p(x|x',\xi)}{p(a'|x')}p(a|x)p(\xi)K_{\xi}\rho_{a|x}K_{\xi}^{\dag}\right] \\
    &= \max_{x'} \sum_{a'} p(a'|x') \Cd_{\text{d}}\left[\sum_{a,x}  p(x)p(a|x) \sum_{\xi} p(\xi) \Xi_{\xi}(\rho_{a|x})\right] \\
    &\leq \max_{x'} \sum_{a'} p(a'|x')\sum_{a,x}  p(x)p(a|x) \Cd_{\text{d}}\left(\sum_{\xi}p(\xi) \Xi_{\xi}(\rho_{a|x})\right) \\
    &\leq \sum_{x} p(x) \sum_{a}p(a|x) \Cd_{\text{d}}\left(\sum_{\xi}p(\xi)  \rho_{a|x}\right) \\
    &\leq \max_{x} \sum_{a}p(a|x) \Cd_{\text{d}}\left( \rho_{a|x}\right) \\
    &= \CCd_{\text{d}}(\mathcal{A}).\label{proof: non-increasing Cd}
\end{aligned}
\end{equation}
Again, in the third line, we utilize the relation: $p(a'|a,x,x',\xi)=p(a'|x',\xi)p(x|a',x',\xi)p(a|x,a',x',\xi)/[p(x|x',\xi)p(a|x,x',\xi)]$.
For the fourth line, the label $x$ should not depend on $a$, and the label $a$ should not depend on $x'$ and $a'$, as shown in Fig.~\ref{fig: LQICC Map}.
In the fifth line, we use the relation $p(x)=p(x|x',\xi)p(a'|a,x,x',\xi)/p(a'|x')$.
In addition, we use the convexity of $\Cd_{\text{d}}$ and its monotonic property under incoherent operations to deduce the inequalities in the sixth and seventh lines, respectively.

In contrast, the $\CH$ will always increase monotonically after the process:
\begin{equation}
\begin{aligned}
    \CH\left[\Xi(\mathcal{A})\right] 
    &= \min_{x'} \sum_{a'} p(a'|x') \Hd \left[\frac{\sigma_{a'|x'}}{p(a'|x')}\right] \\
    &= \min_{x'} \sum_{a'} p(a'|x') \Hd\left[\sum_{a,x,\xi}  p(x)p(a|x)p(\xi)\Xi_{\xi}(\rho_{a|x})\right] \\
    &\geq \sum_{a,x}  p(x)p(a|x) \Hd\left[\sum_{\xi}p(\xi) \Xi_{\xi}(\rho_{a|x})\right] \\
    &= \sum_{x}  p(x) \sum_{a}p(a|x) \Hd\left[\sum_{\xi} p(\xi)\left( \sum_{k} K_{k,\xi}\rho_{a|x} K^{\dag}_{k,\xi}\right)\right] \\
    % &= \sum_{x} p(x) \sum_{a}p(a|x) \Hd \left[  \sum_{i,j,k,\xi} p(\xi) c^{k,\xi}_{i}\ket{i}\bra{i} \rho_{a|x}c^{k,\xi *}_{j}\ket{j}\bra{j} \right] \\
    &= \sum_{x} p(x) \sum_{a}p(a|x) S\left[  \sum_{i,k,\xi} p(\xi) |c^{k,\xi}_{i}|^{2}  \bra{i} \rho_{a|x} \ket{i} \ket{i} \bra{i} \right] \\
    &= \sum_{x} p(x) \sum_{a}p(a|x) S\left[  \Delta(\rho_{a|x})  \right] \\
    &\geq \min_{x} \sum_{a}p(a|x) \Hd\left( \rho_{a|x}\right) \\
    &= \CH(\mathcal{A}).\label{proof: non-increasing CH}
\end{aligned}
\end{equation}

% \red{
% \begin{equation}
% \begin{aligned}
%     \CH\left[\Xi(\mathcal{A})\right] 
%     &= \min_{x'} \sum_{a'} p(a'|x') \Hd \left[\frac{\sigma_{a'|x'}}{p(a'|x')}\right] \\
%     &= \min_{x'} \sum_{a'} p(a'|x') \Hd\left[\sum_{a,x,\xi}  p(x)p(a|x)p(\xi)\Xi_{\xi}(\rho_{a|x})\right] \\
%     &\geq \sum_{a,x}  p(x)p(a|x) \Hd\left[\sum_{\xi}p(\xi) \Xi_{\xi}(\rho_{a|x})\right] \\
%     &= \sum_{x}  p(x) \sum_{a}p(a|x) S\left[\sum_{\xi} p(\xi)\left( \sum_{k} K_{k,\xi} \Delta(\rho_{a|x}) K^{\dag}_{k,\xi}\right)\right] \\
%     &= \sum_{x} p(x) \sum_{a}p(a|x) S\left[  \sum_{i,k,\xi} p(\xi) |c^{k,\xi}_{i}|^{2}  \bra{i} \rho_{a|x} \ket{i} \ket{\Pi_{k}(i)} \bra{\Pi_{k}(i)} \right] \\
%     &= \sum_{x} p(x) \sum_{a}p(a|x) S\left[  \Delta(\rho_{a|x})  \right] \\
%     &\geq \min_{x} \sum_{a}p(a|x) \Hd\left( \rho_{a|x}\right) \\
%     &= \CH(\mathcal{A}).\label{proof: non-increasing CH}
% \end{aligned}
% \end{equation}}

In this derivation, we use the concavity of the $H_\Delta$ and the definition of genuinely incoherent operations in Eq.~\eqref{eq: gico}. 
Therefore, by using a relation similar to Eq.~\eqref{proof: monotonic relation}, we can conclude that $\mathcal{V}_{\text{S}}(\mathcal{A})\geq\mathcal{V}_{\text{S}}\left[\Xi(\mathcal{A})\right]$, which ends the proof.
\end{proof}

One can observe that our proof strategy aims to demonstrate:
\begin{equation}
    \CCd_{\text{d}}(\mathcal{A}) - \CCd_{\text{d}}\left[\Xi(\mathcal{A})\right] \geq 0 \geq \CH(\mathcal{A}) - \CH\left[\Xi(\mathcal{A})\right],\label{eq: former}
\end{equation}
which offers a weaker validation of the relationship 
\begin{equation}
    \CCd_{\text{d}}(\mathcal{A}) - \CH(\mathcal{A})  \geq \CCd_{\text{d}}\left[\Xi(\mathcal{A})\right] - \CH\left[\Xi(\mathcal{A})\right].\label{eq: latter}
\end{equation}
This is because any $\Xi$ satisfying Eq.~\eqref{eq: former} automatically meets the conditions of Eq.~\eqref{eq: latter}. 
However, the converse is not necessarily true; that is, all $\Xi$ that meet the conditions of Eq.~\eqref{eq: latter} may not satisfy Eq.~\eqref{eq: former}. 
To derive the former inequality, we must limit the local operations to genuinely incoherent operations, which form a subset of incoherent operations.

\subsection*{Numerical evidence for the monotonicity of the SIVP}
We have numerically tested the monotonicity of the SIVP using $10^7$ random pure entangled states $\rho^{AB}$ and random local incoherent maps $\Lambda(\rho)=\sum_{\mu} K_{\mu}\rho K_{\mu}^{\dag}$ on Bob's side by setting:
\begin{equation}
    K_{\mu} = c_{\mu} \ket{f_{\mu}(i)}\bra{i} \quad \text{s.t.}~~K_{\mu} \rho K_{\mu}^{\dag} \in \mathcal{I} \quad \forall \rho \in \mathcal{I}\quad \text{and}  \quad \sum_{\mu} K_{\mu}^{\dag} K_{\mu} = \id,
\end{equation}
where $f_{\mu}(i)$ are arbitrary functions used to map one reference basis $\ket{i}$ into another reference basis $\ket{f_{\mu}(i)}$.
For each pure entangled state, we assume that Alice performs the Pauli X ($\sigma_{1}$) and Z ($\sigma_{3}$) measurements, i.e., $M_{a|x} = [\id + (-1)^{a} \sigma_{x} ]/2$, with $a \in \{0,1 \}$ and $x \in \{1,3 \}$.
After receiving the assemblage $\mathcal{A} = \{\sigma_{a|x}=p(a|x)\rho_{a|x}\}_{a,x}$, Bob computes both $\mathcal{V}_{\text{S}}(\mathcal{A})$ and $\mathcal{V}_{\text{S}}[\Lambda(\mathcal{A})]$ under the reference basis $\{\ket{i}\}_{i=0,1}$ (eigenbasis of Pauli-Z).
% Out of all $10^7$ random tests, we found $14$ cases where the SIVP increased after applying the random CPTP maps. 
% One of these $14$ cases is illustrated below:
Out of all $10^7$ random tests, we found that the SIVP always decreased after local incoherent maps. 
Therefore, we conclude from the numerical tests that the SIVP could be non-increasing under one-way local incoherent operations and classical communication~\cite{Chitambar2016PRL}.

Local CPTP maps, on the other hand, could increase the SIVP. 
Given that some CPTP maps may generate local coherence and are not belong to incoherent operations.
For instance, consider the following state $\rho^{AB}$ as
\begin{equation}
    \rho^{AB} = \left(\begin{matrix}
    0.276 & 0.293 - 0.062i & -0.027 + 0.251i & 0.073 - 0.203i \\ 
    0.293 + 0.062i & 0.325 & -0.085 + 0.026i & 0.123 - 0.199i \\
    -0.027 - 0.251i & -0.085 - 0.026i & 0.230 & -0.191 - 0.047i \\
    0.073 + 0.203i & 0.123 + 0.199i & -0.191 + 0.047i & 0.168 
    \end{matrix}\right)
\end{equation}
and a CPTP map: $\Lambda_{1}(\rho) = \sum_{i=0}^{3} K_{i}\rho K_{i}^{\dag}$ with the Kraus operators:
\begin{equation}
\begin{aligned}
     &K_{0} = \left(\begin{matrix}
     0.559 + 0.351i & 0.425 - 0.487i  \\ 
     0.721 & -0.024 + 0.564i
    \end{matrix}\right), \quad
    K_{1} = \left(\begin{matrix}
     0.004 +0.021i & 0.388  \\ 
     -0.160 - 0.030i &  0.319 - 0.091i
    \end{matrix}\right), \\
    &K_{2} = \left(\begin{matrix}
     -0.050 - 0.071i & 0.032 + 0.020i  \\ 
     0.097 &  0.005-0.037i
    \end{matrix}\right), \quad
    K_{3} = \left(\begin{matrix}
    0.021 & 0.006 + 0.012i  \\ 
    0.001 -0.012i & -0.013 - 0.016i 
    \end{matrix}\right).
\end{aligned}
\end{equation}
By calculating the SIVPs, we obtain
\begin{equation}
    \mathcal{V}_{\text{S}}(\mathcal{A})\approx 0.061 \quad \text{and} \quad \mathcal{V}_{\text{S}}[\Lambda_{1}(\mathcal{A})] \approx 0.198,
\end{equation}
which demonstrates that the SIVP increases after the CPTP map. 
However, when applying this map to a maximally mixed state $\id/2$, we find that
\begin{equation}
    \Lambda_{1}\left(\frac{\id}{2}\right) = \left( \begin{matrix}
        0.506 & 0.117 + 0.026i \\
        0.117 - 0.026i & 0.494
    \end{matrix} \right),
\end{equation}
which implies that $\Lambda_{1} \notin$ ICO.
In fact, $\Lambda_{1}[\rho_{\text{I}}(\alpha)]\notin$ ICO $\forall \rho_{\text{I}} = \alpha \ket{0}\bra{0} + (1-\alpha) \ket{1}\bra{1}~\in \mathcal{I} ~\forall \alpha \in [0,1]$.

\subsection*{Proof that SIVP serves as an incompatibility monotone}\label{apx: Proof of property 4}
$\mathcal{V}_{\text{I}}(\mathcal{M})$ is a valid incompatibility monotone~\cite{Paul2019PRL} if it satisfies:
\begin{itemize}
    \item[(a)] $\mathcal{V}_{\text{I}}(\mathcal{M})=0$, if $\mathcal{M}$ is jointly measurable.
    \item[(b)] $\mathcal{V}_{\text{I}}(\mathcal{M})$ satisfies convexity.
    \item[(c)] $\mathcal{V}_{\text{I}}(\mathcal{M})$ is non-increasing under post-processing, namely
    \begin{equation}
    \{M_{a'|x'}\}_{a',x'}= \mathcal{W}(\{M_{a|x}\}_{a,x}) = \sum_{a,x}p(x|x')p(a'|a,x,x')\{M_{a|x}\}_{a,x}.\label{eq: wiring map for measurements}
    \end{equation}
\end{itemize}

\subsubsection*{Proof of condition (a):}
The condition (a) is automatically satisfied by the definition of $\mathcal{V}_{\text{I}}$:
\begin{equation}
    \mathcal{V}_{\text{I}}(\{M_{a|x}\}_{a,x}) = \sup_{\rho_B} \mathcal{V}_{\text{S}}[\sqrt{\rho_B} \{M_{a|x}\}_{a,x} \sqrt{\rho_B}].
\end{equation}

Given that a set of measurements $\{M_{a|x}\}_{a,x}$ is compatible if and only if its steering-equivalent observables induced a state assemblage $\sqrt{\rho_B} \{M_{a|x}\}_{a,x} \sqrt{\rho_B}$ that can be described by an LHS model.
Thus, the SIVP vanishes.
We note that when the shared state is pure entangled, i.e., $\ket{\psi^{AB}}=\sum_i \sqrt{q_i} \ket{i}\ket{i}$, the SEOs $\{\mathcal{B}_{a|x}\}_{a,x}$ is equivalent to the transpose of Alice's measurement $\{M_{a|x}\}_{a,x}$, which is
\begin{equation}
\begin{aligned}
    \mathcal{B}_{a|x} 
    &= \rho_{B}^{-1/2}  \sigma_{a|x} \rho_{B}^{-1/2} \\
    &= \rho_{B}^{-1/2} \left[ \text{Tr}_{A} ~(M_{a|x}\otimes \mathbb{1} )\sum_{i,j}\sqrt{q_iq_j} \ket{i}\bra{j} \otimes \ket{i}\bra{j} \right] \rho_{B}^{-1/2}\\
    &=  \rho_{B}^{-1/2} \sum_{i,j} \sqrt{ q_i q_j} \bra{i} M_{a|x}^{T} \ket{j}  \ket{i}\bra{j} \rho_{B}^{-1/2}  \\
    &=  M_{a|x}^{T}.
\end{aligned}
\end{equation}

\subsubsection*{Proof of condition (b):}
To prove that (b) $\mathcal{V}_{\text{I}}(\mathcal{M})$ satisfies convexity, we only need to demonstrate $\mathcal{V}_{\text{S}}({\mathcal{A}})$ is convex in assemblage.

Let us consider a convex combination of the state assemblages that can be described by 
\begin{equation}
\tilde{\mathcal{A}}= q\mathcal{A}+(1-q)\mathcal{A}':=\{q\sigma_{a|x}+(1-q)\sigma'_{a|x}\}_{a,x}\quad \forall q \in [0,1].    
\end{equation}
Using the convexity of $\Cd_{\text{d}}$, one can obtain
\begin{equation}
\begin{aligned}
    &\CCd_{\text{d}}(\tilde{\mathcal{A}}) = \CCd_{\text{d}}\left[ q\mathcal{A} + (1-q)\mathcal{A}' \right] \\
    &= \max_{x}   \sum_a p(a|x) \Cd_{\text{d}}\left[ q\rho_{a|x} + (1-q)\rho'_{a|x}\right] \\
    &\leq \max_{x} \sum_a \left[ q p(a|x)\Cd_{\text{d}}(\rho_{a|x}) + (1-q)p'(a|x)\Cd_{\text{d}}(\rho'_{a|x}) \right] \\ 
    &\leq q \max_{x}  \sum_a p(a|x)\Cd_{\text{d}}(\rho_{a|x}) 
     + (1-q) \max_{x} \sum_a p'(a|x)\Cd_{\text{d}}(\rho'_{a|x}) \\
    &= q \CCd_{\text{d}}(\mathcal{A}) + (1-q) \CCd_{\text{d}}(\mathcal{A}').
\end{aligned}
\end{equation}

Following the analogous steps, together with the fact that the Shannon entropy is concave, we can demonstrate that $\CH$ is also concave with respect to a convex combination of the state assemblages.
Therefore, we can conclude that the steering violation satisfies convexity, namely 
\begin{equation}
\begin{aligned}
    &\mathcal{V}_{\text{S}}(\tilde{\mathcal{A}})= \mathcal{V}_{\text{S}}\left[ q\mathcal{A}+(1-q)\mathcal{A}' \right] \\
    &= \max \left\{ \CCd_{\text{d}}[q\mathcal{A}+(1-q)\mathcal{A}'] - \CH[q\mathcal{A}+(1-q)\mathcal{A}'],0 \right\} \\
    &\leq \max \{  q\left[ \CCd_{\text{d}}(\mathcal{A}) - \CH(\mathcal{A}) \right] 
     + (1-q)\left[ \CCd_{\text{d}}(\mathcal{A}') - \CH(\mathcal{A}') \right] ,0\} \\
    &\leq q\max \left\{  \ \CCd_{\text{d}}(\mathcal{A}) - \CH (\mathcal{A}) ,0\right\} 
     + (1-q) \max \left\{  \CCd_{\text{d}}(\mathcal{A}') - \CH (\mathcal{A}'),0 \right\} \\
    &=q\mathcal{V}_{\text{S}}(\mathcal{A}) + (1-q)\mathcal{V}_{\text{S}}(\mathcal{A}').\label{eq: convexity}
\end{aligned}
\end{equation}
Here, we use the facts that $\Cd_{\text{d}}$ ($\CH$) is a convex (concave) function and the property of the maximization in order.
Therefore, we conclude the proof that $\mathcal{V}_{\text{I}}(\mathcal{M})$ satisfies convexity.

\subsubsection*{Proof of condition (c):}
\begin{figure}[!htbp]
    \centering
    \includegraphics[width=0.5\columnwidth]{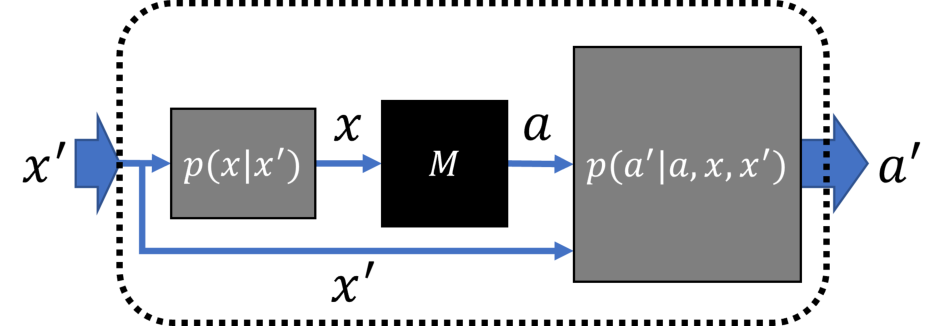}
    \caption{Schemetic illustration of post-processing. See the caption in Fig.~\ref{fig: LQICC Map}}\label{fig: post-processing}
\end{figure}

To prove that $\mathcal{V}_{\text{I}}(\mathcal{M})$ is non-increasing under post-processing, we consider a post-processing scenario $\mathcal{W}$ defined as a deterministic wiring map~\cite{Gallego2015PRX}, as shown in Fig.~\ref{fig: post-processing}:
\begin{equation}
    \sigma_{a'|x'} = \mathcal{W}(\sigma_{a|x}) = \sum_{a,x}p(x|x')p(a'|a,x,x')\sigma_{a|x},~~\forall a,x, \label{eq: wiring map}
\end{equation}
where $p(x|x')$ and $p(a'|a,x,x')$ are conditional probabilities.
We first need to demonstrate that $\mathcal{V}_{\text{S}}(\mathcal{A})$ is also non-increasing under post-processing.
We begin this proof by showing that $\CCd_{\text{d}}$ is non-increasing under post-processing, that is
\begin{equation}
\begin{aligned}
    \CCd_{\text{d}}\left[\mathcal{W}(\mathcal{A})\right]
    &= \max_{x'} \sum_{a'} p(a'|x') \Cd_{\text{d}} \left[\frac{\sigma_{a'|x'}}{p(a'|x')}\right] \\
    &= \max_{x'} \sum_{a'} p(a'|x') \Cd_{\text{d}}\left[\sum_{a,x} \frac{  p(x|x')p(a'|a,x,x')\sigma_{a|x} }{p(a'|x')}\right] \\
    &= \max_{x'} \sum_{a'} p(a'|x') \Cd_{\text{d}}\left[\sum_{a,x} \frac{ p(x|x')}{p(a'|x')} \frac{p(a'|x')p(x|x',a')p(a|x,x',a')}{p(x|x')p(a|x,x')}p(a|x)\rho_{a|x}\right] \\
    &= \max_{x'} \sum_{a'} p(a'|x') \Cd_{\text{d}}\left[\sum_{a,x} \frac{ p(x|x')}{p(a'|x')} \frac{p(a'|x')p(x|x')p(a|x,x')}{p(x|x')p(a|x,x')}p(a|x)\rho_{a|x}\right] \\
    &= \max_{x'} \sum_{a'} p(a'|x') \Cd_{\text{d}}\left[\sum_{a,x} \frac{p(a'|x')p(x|x')}{p(a'|x')}p(a|x)\rho_{a|x}\right] \\
    &= \max_{x'} \sum_{a'} p(a'|x') \Cd_{\text{d}}\left[\sum_{a,x}  p(x)p(a|x) \rho_{a|x}\right] \\
    &\leq \max_{x'} \sum_{a'} p(a'|x')\sum_{a,x}  p(x)p(a|x) \Cd_{\text{d}}\left(\rho_{a|x}\right) \\
    &= \sum_{x} p(x) \sum_{a}p(a|x) \Cd_{\text{d}}\left( \rho_{a|x}\right) \\
    &\leq \max_{x} \sum_{a}p(a|x) \Cd_{\text{d}}\left( \rho_{a|x}\right) \\
    &= \CCd_{\text{d}}(\mathcal{A}).\label{proof: post-process monotone Cd}
\end{aligned}
\end{equation}
Here, we utilize the relation $p(a'|a,x,x') =p(a'|x')p(x|x',a')p(a|x,x',a')/[p(x|x')p(a|x,x')]$ to arrive at the equation in the third line; in the fourth line, we note that all the labels, say $x,x',a$ should not dependent on $a'$; in the sixth line, we use the relation of $p(x|x')=p(x)p(x'|x)/p(x')=p(x)$, given that $x'$ should not depend on $x$; the seventh line uses the convexity of $C_{\text{d}}$.

In contrast, the conditional Shannon entropy increases monotonically after the process due to concavity, namely
\begin{equation}
\begin{aligned}
    \CH\left[\mathcal{W}(\mathcal{A})\right] 
    &= \min_{x'} \sum_{a'} p(a'|x') \Hd \left[\frac{\sigma_{a'|x'}}{p(a'|x')}\right] \\
    &= \min_{x'} \sum_{a'} p(a'|x') \Hd\left[\sum_{a,x}  p(x)p(a|x)\rho_{a|x}\right] \\
    &\geq \sum_{a,x}  p(x)p(a|x) \Hd\left(\rho_{a|x}\right) \\
    &\geq \min_{x} \sum_{a}p(a|x) \Hd\left( \rho_{a|x}\right) \\
    &= \CH(\mathcal{A}).\label{proof: post-process monotone CH}
\end{aligned}
\end{equation}
Combining the results in Eqs.~\eqref{proof: post-process monotone Cd} and \eqref{proof: post-process monotone CH}, one can conclude that 
\begin{equation}
    \CCd_{\text{d}}(\mathcal{A}) - \CCd_{\text{d}}\left[\mathcal{W}(\mathcal{A})\right] \geq 0 \geq \CH(\mathcal{A}) - \CH\left[\mathcal{W}(\mathcal{A})\right],\label{proof: monotonic relation}
\end{equation}
which implies $\mathcal{V}_{\text{S}}(\mathcal{A})\geq\mathcal{V}_{\text{S}}\left[\mathcal{W}(\mathcal{A})\right] $.

By using the above results, we can therefore show that $\mathcal{V}_{\text{I}}(\mathcal{M})$ is also non-increasing under post-processing, i.e.,
\begin{equation}
\begin{aligned}
    \mathcal{V}_{\text{I}}\left[\mathcal{W}(\mathcal{M})\right] 
    &= \sup_{\rho_{B}} \mathcal{V}_{\text{S}} \left[ \sqrt{\rho_{B}} \mathcal{W}(\mathcal{M})\sqrt{\rho_{B}}\right] \\
    &= \mathcal{V}_{\text{S}} \left[ \sqrt{\rho^{\star}_{B}} \mathcal{W}(\mathcal{M})\sqrt{\rho^{\star}_{B}}\right] \\
    &= \mathcal{V}_{\text{S}} \left[ \mathcal{W}\left(\sqrt{\rho^{\star}_{B}} \mathcal{M}\sqrt{\rho^{\star}_{B}}\right)\right] \\
    &\leq \mathcal{V}_{\text{S}}\left(\sqrt{\rho^{\star}_{B}} \mathcal{M}\sqrt{\rho^{\star}_{B}}\right) \\
    &\leq \sup_{\rho_{B}}\mathcal{V}_{\text{S}} \left(\sqrt{\rho_{B}} \mathcal{M}\sqrt{\rho_{B}}\right) \\
    &= \mathcal{V}_{\text{I}}(\mathcal{M}). 
\end{aligned}
\end{equation}

\section*{Data availability}
Dataset underlying the results presented in Fig. 2 is available in Ref.~\cite{lee2024dataset}.

\section*{Code availability}
Source codes of the plots are available from the authors upon request.

\section*{Acknowledgments}
We are grateful to Eric Chitambar for fruitful discussion. 
A.M. is supported by the Polish National Science Centre (NCN) under the Maestro Grant No. DEC-2019/34/A/ST2/00081.
A.Č. and K.L. acknowledge support by the project OP JAC
CZ.02.01.01/00/22\_008/0004596 of the Ministry of Education, Youth, and Sports of the Czech Republic and EU.
F.N. is supported in part by: the Japan Science and Technology Agency (JST) [via the CREST Quantum Frontiers program Grant No. JPMJCR24I2, the
Quantum Leap Flagship Program (Q-LEAP), and the Moonshot R\&D Grant Number JPMJMS2061], and the Office of Naval Research (ONR) Global (via Grant No. N62909-23-1-2074).
H.-Y.~K. is supported by the
Ministry of Science and Technology, Taiwan, (Grants No.~MOST 111-2917-I-564-005, 112-2112-M-003 -020 -MY3), and Higher Education Sprout Project of National Taiwan Normal
University (NTNU) and the Ministry of Education (MOE) in Taiwan.
This work is supported by the National Center for Theoretical Sciences and National Science and Technology Council, Taiwan, Grants No. NSTC 113-2123-M-006-001.

\section*{Author contributions}

K.-Y. L. and J.-D. L. contributed equally to this work.
K.-Y. L. and J.-D. L. performed the computations and proved the properties with H.-Y. K.'s help. 
K. L. and A. \v{C} designed the experimental setup and conducted the experiment; they analyzed the experimental data together with A. M. and K.-Y. L.
K.-Y. L., J.-D. L., K. L., and H.-Y. K. contributed to analyzing the results and wrote the first draft of the manuscript.
F. N., H.-Y. K. and Y.-N. C. supervised the research and were responsible for the integration among different research units. 
All authors contributed to the discussion of the central ideas and reviewed the manuscript.

\section*{Competing interests}

The authors declare no competing interests.

% \bibliography{RefPrl.bib}

\end{document}